\documentclass[aps,preprint,onecolumn,floatfix,superscriptaddress,nofootinbib]{revtex4-1}
\newcommand{\sigsip}{\ensuremath{\sigma_{\chi p}}}
\newcommand{\kev}{\ensuremath{\,\mathrm{keV}}}
\newcommand{\mev}{\ensuremath{\,\mathrm{MeV}}}
\newcommand{\gev}{\ensuremath{\,\mathrm{GeV}}}
\newcommand{\tev}{\ensuremath{\,\mathrm{TeV}}}

\usepackage{amsmath}
\usepackage{graphicx,bm}
\usepackage[colorlinks,citecolor=blue]{hyperref}
\usepackage[caption=false]{subfig}
\usepackage{slashed}

\begin{document}
\title{Revising inelastic dark matter direct detection by including the cosmic ray acceleration}

\author{Jie-Cheng Feng}
\affiliation{Department of Physics, Beijing Normal University, Beijing 100875, China}

\author{Xian-Wei Kang}
\email{xwkang@bnu.edu.cn}
\affiliation{Key Laboratory of Beam Technology of the Ministry of Education, 
College of Nuclear Science and Technology, Beijing Normal University, Beijing 100875, China}
\affiliation{Beijing Radiation Center, Beijing 100875, China}

\author{Chih-Ting Lu}
\affiliation{School of Physics, KIAS, Seoul 130-722, Republic of Korea}

\author{\\Yue-Lin Sming Tsai}
\email{smingtsai@pmo.ac.cn}
\affiliation{Key Laboratory of Dark Matter and Space Astronomy, Purple Mountain Observatory, Chinese Academy of Sciences, Nanjing 210033, China}

\author{Feng-Shou Zhang}
\affiliation{Key Laboratory of Beam Technology of the Ministry of Education, 
College of Nuclear Science and Technology, Beijing Normal University, Beijing 100875, China}
\affiliation{Beijing Radiation Center, Beijing 100875, China}

\begin{abstract}
The null signal from collider and dark matter (DM) direct detector experiments makes the interaction between DM and visible matter too small to reproduce the correct relic density for many thermal DM models.
The remaining parameter space indicates that
two almost degenerated states in the dark sector, the inelastic DM scenario, can co-annihilate in the early universe to
produce the correct relic density. Regarding the direct detection of the inelastic DM scenario,
the virialized DM component from the nearby halo is nonrelativistic
and not able to excite the DM ground state, even if the relevant couplings can be considerable.
Thus, a DM with a large mass splitting can evade traditional virialized DM direct detection.
In this study, we connect the concept of cosmic-ray accelerated DM in our Milky Way and
the direct detection of inelastic scattering in underground detectors
to explore spectra that result from several interaction types of the inelastic DM.
We find that the mass splitting $\delta<\mathcal{O}(1\gev)$ can still be reachable for cosmic ray accelerated DM with mass range $1\mev<m_{\chi_1}<100\gev$ and sub-GeV light mediator using the latest PandaX-4T data, even though we conservatively use the astrophysical parameter
(effective length) $D_{\rm eff}=1$~kpc.
\end{abstract}

\date{\today}

\maketitle
%\flushbottom

\section{introduction}

The gravitational evidence of dark matter (DM) is strong and clear. 
However, its nongravitational interaction with the standard model (SM) has not yet been observed.
Conversely, if the DM number density in the early universe
can be described by the thermal Boltzmann distribution, such as the SM particles,
the Planck measured relic density~\cite{Planck:2018jri} implies that
DM must interact with the SM besides gravity.
Among the various methods to detect the interaction between DM and SM, laboratory measurements, including the Large Hadron Collider (LHC)~\cite{ATLAS:2021shl,CMS:2021far} and DM direct detection (DD)~\cite{XENON:2018voc,PandaX:2021osp}, provide the most robust searches.
However, only null signals have been reported.
Particularly, the limits from either XENON1T~\cite{XENON:2018voc} or PandaX-4T~\cite{PandaX:2021osp}
rule out the DM-proton elastic scattering cross section close to
the neutrino floor.
These stringent constraints squeeze the allowed DM model parameter space to some fine-tuning regions
where the correct relic density is generated by some special mechanisms
(e.g., resonance or coannihilation~\cite{Griest:1990kh}).
The coannihilation region indicates that
the lightest DM particle $\chi_1$ and the next lightest one $\chi_2$ almost degenerate in mass.
When the coannihilation mechanism governs DM annihilations,
the $\chi_1-\chi_1-$SM coupling can be negligible.
Thus, the $\chi_1-p$ elastic scattering will be suppressed~\cite{Edsjo:1997bg,Baer:2005jq,Cheung:2012qy,Nagata:2015pra,Banerjee:2016hsk,Tsai:2019eqi}.
In contrast, the resonance region (the mass of the mediator such as Higgs equal to twice the DM mass)
may be completely probed in future DD sensitivities
(e.g., see Higgs portal DM models~\cite{Cheung:2012xb,Cline:2013gha,Lin:2013sca,GAMBIT:2017gge,Matsumoto:2018acr,Arcadi:2019lka}).

The standard DM DD strategy is to detect DM-nucleon interactions by measuring
the recoil energy of DM-nucleon scattering under the following consideration.
When the Earth sweeps the local virialized halo,
the Maxwell-Boltzmann distributed DM, hereafter called virialized DM (vDM),
hits the detector target.
Because the vDM velocity is non-relativistic,
the currently measured nuclei recoil energy range does not cover DM mass lighter than
$\approx 5\gev$ for xenon-type detectors~\cite{LUX:2016ggv,XENON:2018voc,PandaX:2021osp}.
Also, such a standard method may also be blind when searching
the coannihilation region
because the incoming vDM is non-relativistic and
its small kinetic energy cannot excite DM to the next lightest one, 
namely its kinetic energy smaller than the mass difference between $\chi_1$ to $\chi_2$.
%, even if the relevant coupling can be considerable.
Quantitatively, a DM mass heavier than $\mathcal{O}(\tev)$ is needed to detect an excitation 
from $\chi_1$ to $\chi_2$ with the mass splitting $\approx\mathcal{O}(100\kev)$~\cite{PandaX-II:2017zex}.

\begin{figure}[htp]
    \centering
    \includegraphics[width=0.5\textwidth]{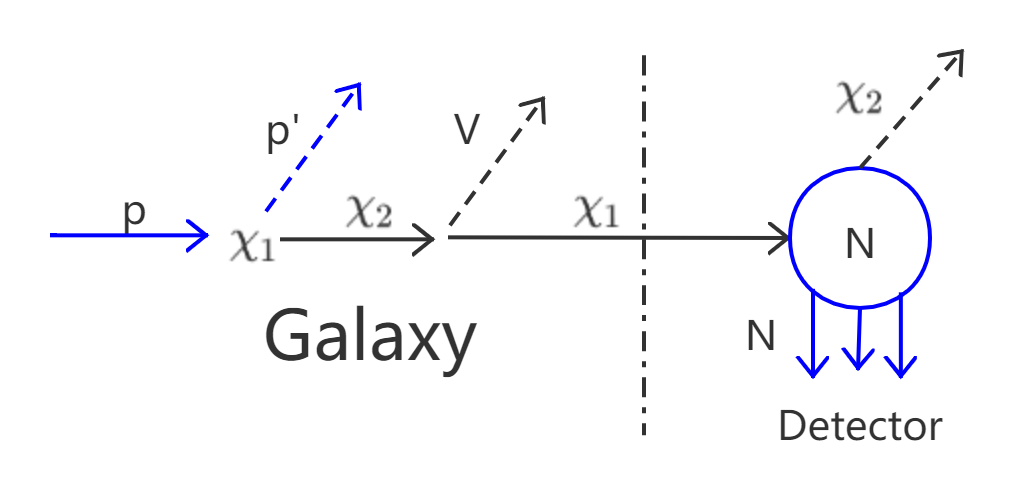}
    \caption{Cosmic ray accelerated inelastic DM detection scenario: $  p+\chi_1 \to p^\prime+\chi_2 \to p^\prime +\chi_1 + V$,
    where $p$ denotes the CR proton or helium, and the prime is used to distinguish the same particles in the initial and final states.
    The inelastic DM $\chi_1$ and $\chi_2$ are the ground state and excited state, respectively; $V$ is the mediator between SM and DM; and
    $N$ is the target nucleus inside the underground detector. By this mechanism, the final $\chi_1$ is accelerated.}
    \label{Fig:scenario}
\end{figure}

Other researchers have proposed searching for accelerated vDMs by considering
their collisions with
the high energy cosmic ray (CR) protons~\cite{Bringmann:2018cvk,Ema:2018bih,Cappiello:2018hsu,Cappiello:2019qsw,Wang:2019jtk}.
When collisions occur, the DM mass below a few $\gev$ can be accelerated to
be relativistic and enter the underground detector with kinetic energy
higher than the designed threshold energy.
This kind of cosmic ray accelerated vDM (CRDM), has recently received much attention 
because it demonstrates what is possible to detect sub-GeV DM with current DM DD experiments. 
Even if CRDM fluxes are several orders of magnitude lower than those of vDM,
the detection of CRDM helps us to probe the DM mass region lighter than a few $\gev$
using the sensitivity of XENONnT and PandaX-4T.
This DM mass region was almost undetectable by using the vDM scenario.
With a similar idea, light CRDM can be detected with a neutrino telescope~\cite{Guo:2020drq}
or by studying the diurnal effect caused by light CRDM~\cite{Ge:2020yuf}.
In addition, such a collision between the CR proton and DM may also smash the proton and
produce neutrino and gamma ray~\cite{Guo:2020oum}.

We extend the CRDM elastic scattering to
the inelastic scattering scenario (see the schematic carton in Fig.~\ref{Fig:scenario}).
The first vertex occurs in the Milky Way, where $\chi_1$ is excited to $\chi_2$ after the first collision.
Unlike the elastic CRDM scenario, $\chi_2$ is not a stable particle and subsequently decays back to $\chi_1$
and emits a mediator particle $V$ before reaching the underground detector.
The velocity of $\chi_1$ here can be higher than elastic CRDM because
the decay from heavier $\chi_2$ can boost $\chi_1$ again.
Recently, investigating a similar scenario with the vector-vector interaction between femionic DM and SM,
Ref.~\cite{Bell:2021xff} found that this approach can probe a larger mass splitting $\approx 100\mev$.
Conversely, the traditional vDM inelastic scattering search is limited to
the heavier mass and low mass splitting~\cite{PandaX-II:2017zex}.

In this study, we consider both elastically and inelastically produced CRDM.
To clarify their difference, we list their productions and mass splitting $\delta$ conditions as follows:
\begin{itemize}
 \item For inelastic scattering, the full process is $p+\chi_1 \to p^\prime+\chi_2 \to p^\prime +\chi_1 + V$
as shown in Fig.~\ref{Fig:scenario}, where the mass splitting $\delta$ between $\chi_1$ and $\chi_2$ is nonzero. 
To simplify the calculation, we assume that $\chi_2$ in the decay is on shell, and as a result, we will focus on the light mediator case in this study, i.e., $\delta>m_V$.
\item The elastically produced CRDM refers to CR-DM elastic scattering $ p+\chi_1 \to p^\prime+\chi_1$. In this situation, $\chi_2$ is not present, and $\delta=0$ as well.
\end{itemize}

We consider that the DM-proton cross section of the vector-vector interaction
is rather a constant, independent of incoming proton energy.
However, the CR spectrum rapidly decreases with CR energy.
Except for the vector-vector interaction between fermionic DM and SM,
there are more possible interaction types, and the velocity-dependent terms can also be important,
which are beyond the popular (velocity-independent) spin-independent (SI)
and spin-dependent (SD) form.
Thus, we study several CRDM spectra based on different types of interactions
to see the impact on the detected event rate.
We investigate the exclusion power of the latest PandaX-4T data~\cite{PandaX:2021osp}
in the region of the lower mass $m_{\chi_1} < 1 \gev$ and
larger mass splitting $\delta \sim \mathcal{O}(\gev)$.

The remainder of this paper is organized as follows.
First, we introduce several different types of DM-SM interactions in Sec.~\ref{sec:models} by
considering both fermionic and scalar DM.
In Sec.~\ref{sec:DMCR}, we describe the detection of elastic and inelastic DM-proton scattering
for both vDM and CRDM.
In Sec.~\ref{sec:pandax4t}, we examine the interactions with the latest and
most stringent exclusion from PandaX-4T.
Finally, we conclude the study in Sec.~\ref{sec:conclusion}.

\section{Effective inelastic DM interactions}
\label{sec:models}

To demonstrate our work, we consider minimum DM Lagrangians
where a $Z_2$ even vector/axial vector mediator $V$\footnote{The leptophobic mediator $V$ will mainly decay to mesons or quarks 
if kinematically allowed. For $m_V < m_{\pi}$, $V$ can still decay to $e^+e^-$, $\nu\overline{\nu}$ via one loop process or $V-Z$ boson mixing. 
Note that the lifetime of $V$ must be shorter than 1 second to avoid spoiling typical Big Bang nucleosynthesis history.}
and a DM $\chi_1$ with its excited state $\chi_2$ are considered
as the new implementation to the SM. We discuss both Majorana and real scalar DM fields in this study.
The common nucleon-$V$ interactions are:
\begin{eqnarray}
\mathcal{L}_{VN} &=& \overline{N}\gamma^{\mu}\left( C^v_{N} + C^a_{N}\gamma^5 \right)N V_\mu.
\label{eq:LSM}
\end{eqnarray}
The couplings $C^v_{N}$ and $C^a_{N}$ are for vector and axial vector interactions for nucleon.
We require their sizes to be less than unity. Otherwise, the DM interaction with the nucleon would be so strong and be observed by detectors.
The mediator $V$ can be either a photon-like boson charged under the $U(1)$ gauge or
$Z$-like boson charged under $SU(2)$ gauge.
To maintain gauge invariance, Ref.~\cite{Gallagher:2020ajd}
indicates that if the boson $V$ is $SU(2)$ dark $Z$,
the tree-level amplitude squared must be computed
with a unitary gauge~\cite{Ruegg:2003ps,Ivanov:2015tru},
which will be considered in this study for both the vector and axial-vector interactions.

Next, we discuss interesting DM-$V$ interactions $\mathcal{L}_{VD}$ and thus the completed Lagrangian
to describe $p + \chi_1\to p' + \chi_2$ process is $\mathcal{L}_{VN}+\mathcal{L}_{VD}$.
Then, we focus on some characteristic types of Lagrangians for demonstration.

\subsection{Fermionic DM interaction}

\begin{figure*}[htbp]
\begin{centering}
\subfloat[$\mathcal{L}^f_{1}$. ]{
\includegraphics[width=0.49\textwidth]{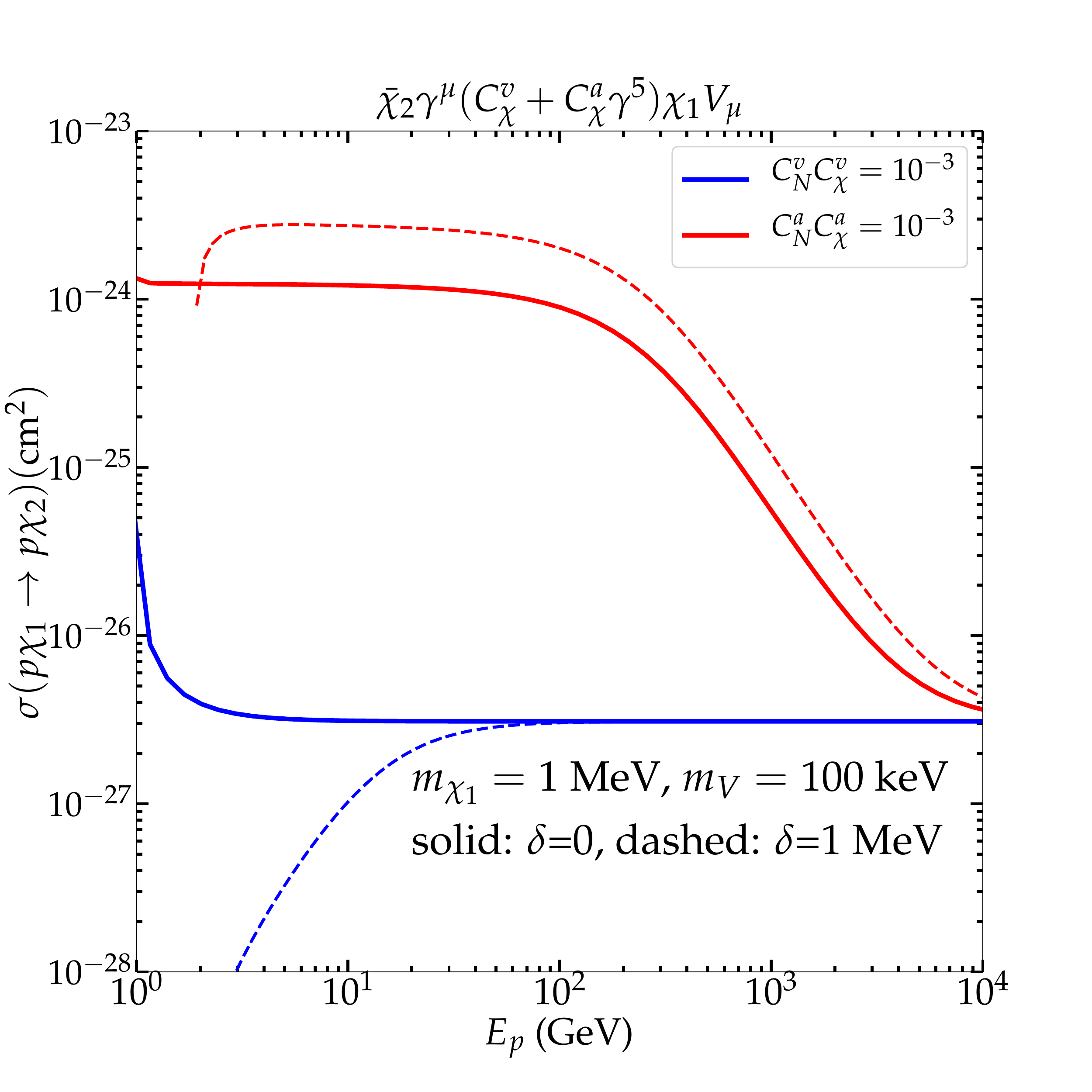}
\label{fig:Ep_Xsec_fer_cacv_a}
}
\subfloat[$\mathcal{L}^f_{2}$]{
\includegraphics[width=0.49\textwidth]{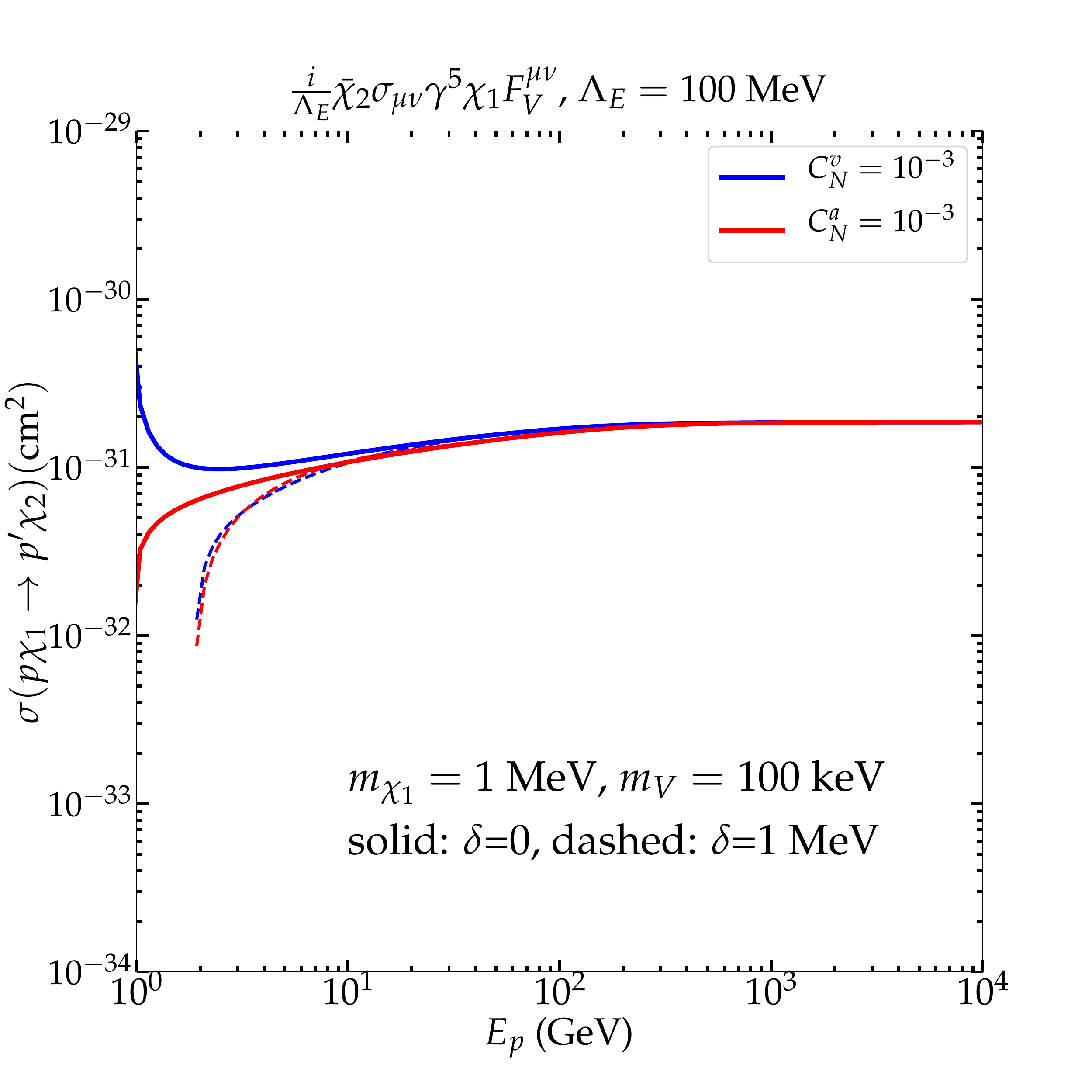}
\label{fig:Ep_Xsec_fer_cacv_b}
}
\caption{
The integrated cross section of process $p \chi_1\to p' \chi_2$ as a function of incoming proton energy $E_p$.
We used red and blue lines to denote $C^v_{N}=0$ and $C^a_{N}=0$, respectively.
Compared with the elastic scenario (solid lines), we also present $\delta=1\mev$ as
dashed lines.
The AA interaction ($C^v_{\chi}=C^v_{N}=0$) has a greater cross section
than VV one ($C^a_{\chi}=C^a_{N}=0$), while
for $\mathcal{L}^f_{2}$, which is illustrated by ED, the resulting difference between the vector interaction
and axial vector interaction of $\mathcal{L}_{VN}$ is small.
}
\label{Fig:Ep_Xsec_fer_cacv}
\end{centering}
\end{figure*}

For the fermionic DM scenario, two common types of effective interactions between DM and $V$ are considered:
\begin{eqnarray}
\mathcal{L}^f_{1} &=& \left[ \overline{\chi_2}\gamma_{\mu}\left( C^v_{\chi} + C^a_{\chi}\gamma^5\right)\chi_1 + h.c. \right] V^{\mu},
\label{eq:f1}\\
\mathcal{L}^f_{2} &=& \frac{1}{\Lambda_{E(M)}} \bar{\chi}_2 \sigma_{\mu\nu}\Gamma_{E(M)}\chi_1 F^{\mu\nu}_V,
\label{eq:f2}
\end{eqnarray}
where $F^{\mu\nu}_V=\partial^\mu V^\nu-\partial^\nu V^\mu$,
$\sigma_{\mu\nu}=\frac{i}{2}(\gamma^\mu\gamma^\nu-\gamma^\nu\gamma^\mu)$,
electric dipole-like interaction $\Gamma_E=i\gamma_5$, and magnetic dipole-like interaction $\Gamma_M=1$.
The index $f$ refers to the Majorana DM.
The parameter $\Lambda_{E(M)}$ is the new physics scale and is a dimensional coupling that cannot be compared with $C^v_{\chi}$
or $C^a_{\chi}$ directly.
Also, we only consider the product of $\chi\chi V$ and $N N V$ couplings
that appears in the scattering amplitude.
Therefore, we can simply fix the coupling constants of the dark sector as constants
$C_{\chi}^{v/a} = 1$ and $\Lambda_{E(M)} = 100~\rm{MeV}$.
We then compare the results based on different values of the $N N V$ coupling, namely, $C^{v}_N$ or $C^{a}_N$, and can easily rescale the result for other choices.

For simplicity, we focus on one type of interaction each time.
In principle, there are four DM-SM interactions for $\mathcal{L}^f_{1}$:
vector-vector (VV), vector-axial vector (VA), axial vector-vector (AV), and axial vector-axial vector (AA).
Similarly, there are four possible DM-SM interactions for $\mathcal{L}^f_{2}$:
vector-magnetic dipole (MD), axial vector-magnetic dipole, vector-electric dipole (ED), and
axial vector-electric dipole.
We choose the representative operators VV, AA, ED, and MD in this study because the VV and AA interaction predicts the minimal and maximum $p-\chi_1$ cross sections, respectively, and
ED and MD are the most familiar DM models in the community.

The cross sections based on Eq.~\eqref{eq:f1} and Eq.~\eqref{eq:f2} are given in Appendix~\ref{app:Xsec2to2}.
For a comparison of axial vector and vector interaction in $\mathcal{L}_{VN}$,
Fig.~\ref{Fig:Ep_Xsec_fer_cacv} shows an integrated cross section of the process
$p \chi_1 \to p' \chi_2$ as a function of incoming proton energy $E_p$ for each interaction of $\mathcal{L}^{f}_{VD}$.
The solid lines represent the degenerate scenario ($m_{\chi_1}=m_{\chi_2}$),
while the dashed lines are based on the inelastic cross section with $\delta=1\mev$.
In Fig.~\ref{fig:Ep_Xsec_fer_cacv_a}, the cross section of AA is higher than that of VV in the lower $E_p$ region 
due to an additional contribution to the cross section of the AA,
which dominates for small values of $E_p$.
When the mediator mass is light, the enhancement is strong.
Mass splitting can enhance axial vector interaction but suppress the vector interaction
at the lower $E_p$ region.
However, in the large $E_p$ region, all the interactions predict almost the same cross section.
For dipole interactions of $\mathcal{L}^{f}_{VD}$, the resulting difference between the vector and axial vector in $\mathcal{L}_{VN}$ is less noticeable, as illustrated in Fig.~\ref{fig:Ep_Xsec_fer_cacv_b}.
Also, when mass splitting $\delta$ exists, there is a kinematic constraint on $E_p$
for the inelastic scattering. More details are shown in Appendix.~\ref{app:kinematics}.
Therefore, the dashed lines do not start at $E_p=m_p$.

\begin{figure*}[htbp]
\begin{centering}
\subfloat[$\mathcal{L}^f_{1}$]{
\includegraphics[width=0.48\textwidth]{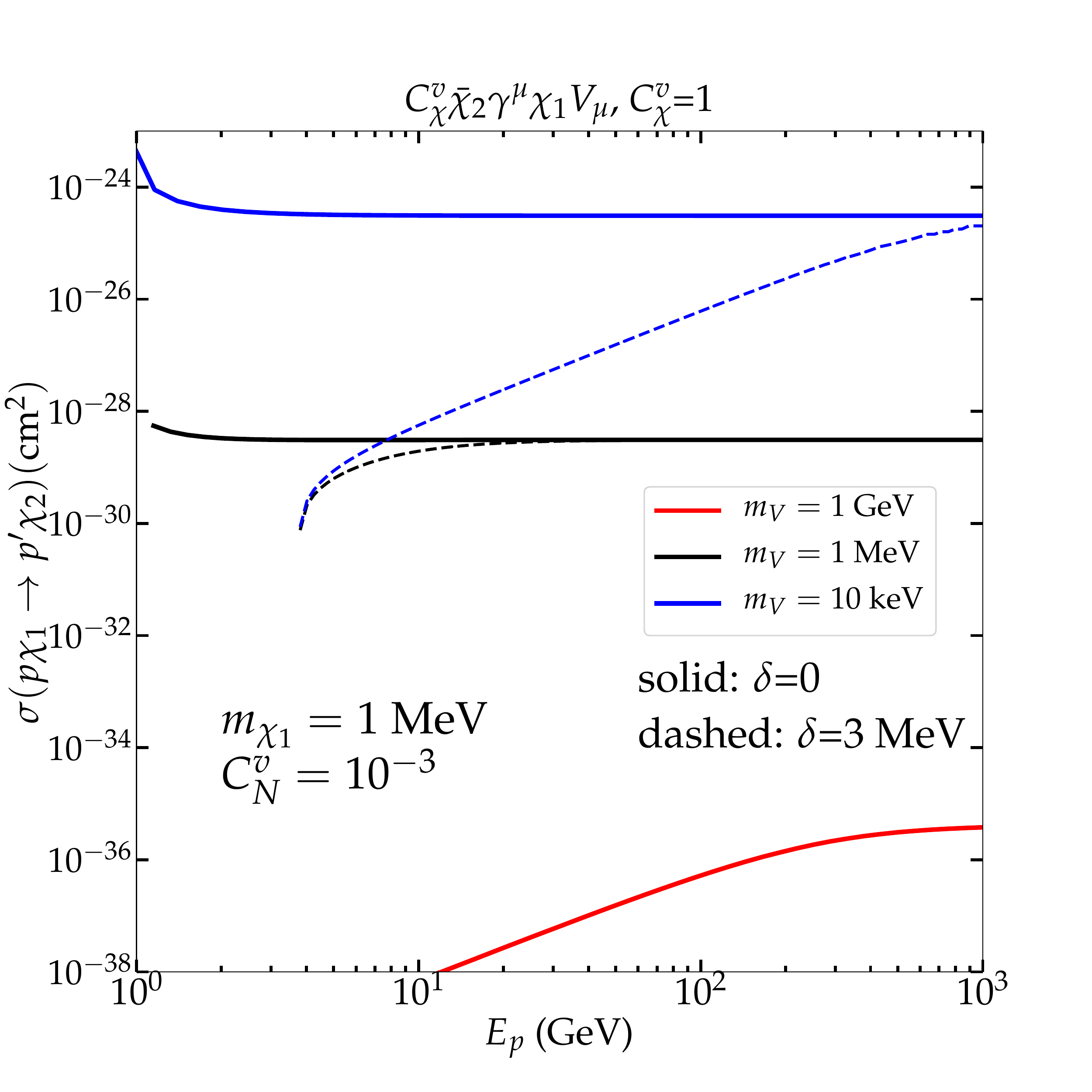}
\label{fig:Ep_Xsec_fer_mv_a}
}
\subfloat[$\mathcal{L}^f_{2}$]{
\includegraphics[width=0.48\textwidth]{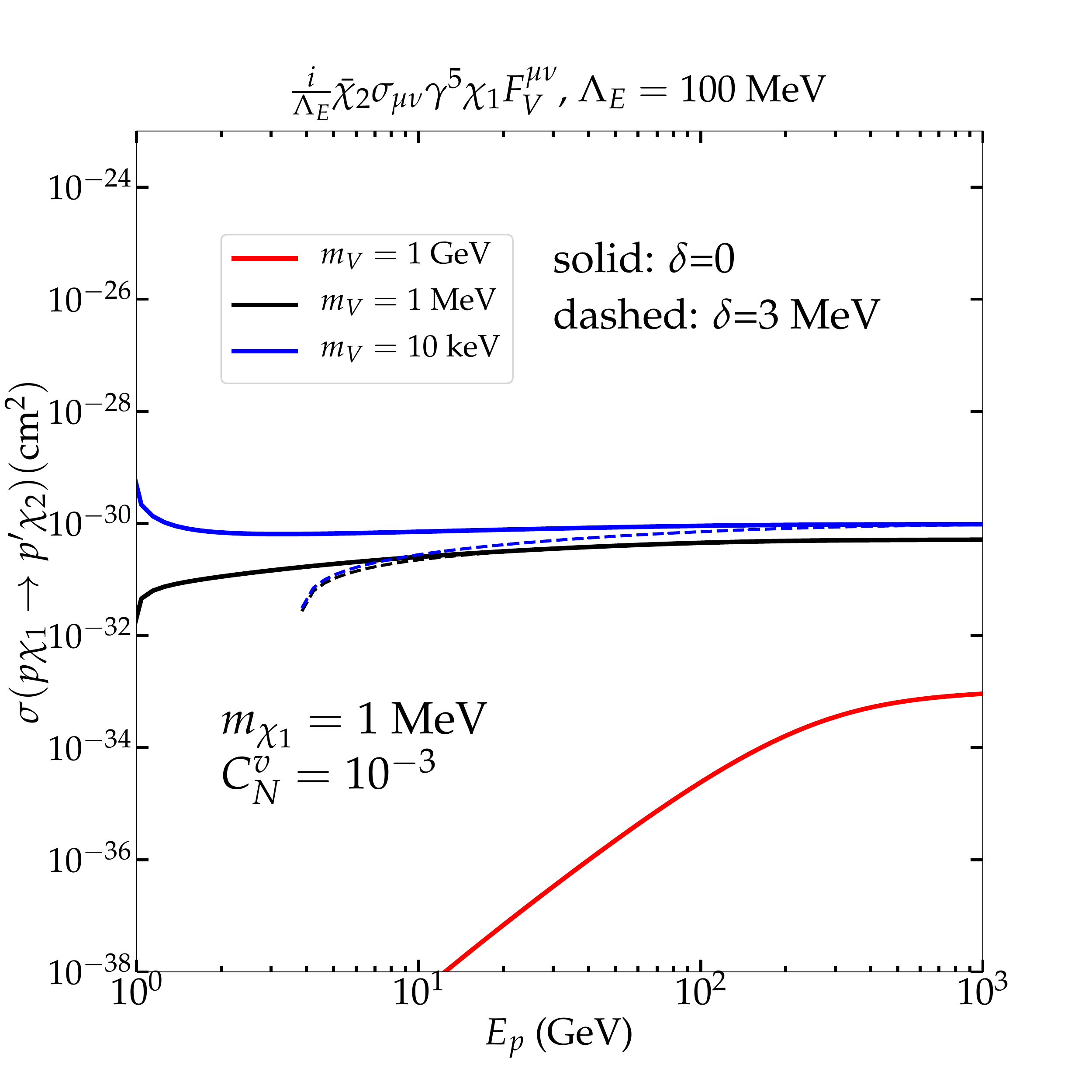}
\label{fig:Ep_Xsec_fer_mv_b}
}
\caption{
The integrated cross section for the process $p\chi_1\to p'\chi_2$ with respect to the energy of incoming proton $E_p$
for $\mathcal{L}^f_{1}$ (left panel) and $\mathcal{L}^f_{2}$ (right panel).
Three mediator masses $10\kev$ (blue lines), $1\mev$ (black lines), and $1\gev$ (red lines) are shown.
The difference between the
red dashed line and red solid line are negligible.
We only plot the SM vector interaction ($C_N^a=0$) case as a demonstration.
}
\label{Fig:Ep_Xsec_fer_mv}
\end{centering}
\end{figure*}

In Fig.~\ref{Fig:Ep_Xsec_fer_mv}, we show the integrated cross sections for $\mathcal{L}^f_{1}$ (left panel) and
$\mathcal{L}^f_{2}$ (right panel) by comparison with three different mediator masses:
$m_V=10\kev$ (blue lines), $m_V=1\mev$ (black lines), and $m_V=1\gev$ (red lines).
As a demonstration, we only present the vector interaction (with $C_N^a=0$) case
while the cross section for the axial vector interaction will be larger, as shown in Fig.~\ref{Fig:Ep_Xsec_fer_cacv}.
Again, we use the solid line for the degenerate scenario and the dashed line for the inelastic scattering.
Generally, both the larger $\delta$ and larger $m_V$ can suppress the cross section at the small
$E_p$ region, but the cross section will eventually be saturated in the high-energy region.
When $\delta\gg m_V$, the cross section can even be suppressed up to the $E_p>100\ \gev$ region (cf. blue dashed line in panel (a)).
Comparing two different fermionic DM interactions, the suppression due to $\delta$ is more severe
in the $\mathcal{L}^f_{1}$ than $\mathcal{L}^f_{2}$ for the light $m_V$ case.

%%%%%%%%%%%%%%%%%%%%%%%%%%%%%%%%%%%%%%%%%%%%%%%%%%%%%%%%%%%%%%%%%%%%%%%%%%%%%
\subsection{Scalar DM interaction}
%%%%%%%%%%%%%%%%%%%%%%%%%%%%%%%%%%%%%%%%%%%%%%%%%%%%%%%%%%%%%%%%%%%%%%%%%%%%%

\begin{figure*}[htbp]
\begin{centering}
\subfloat[$\mathcal{L}^s_{1}$. ]{
\includegraphics[width=0.48\textwidth]{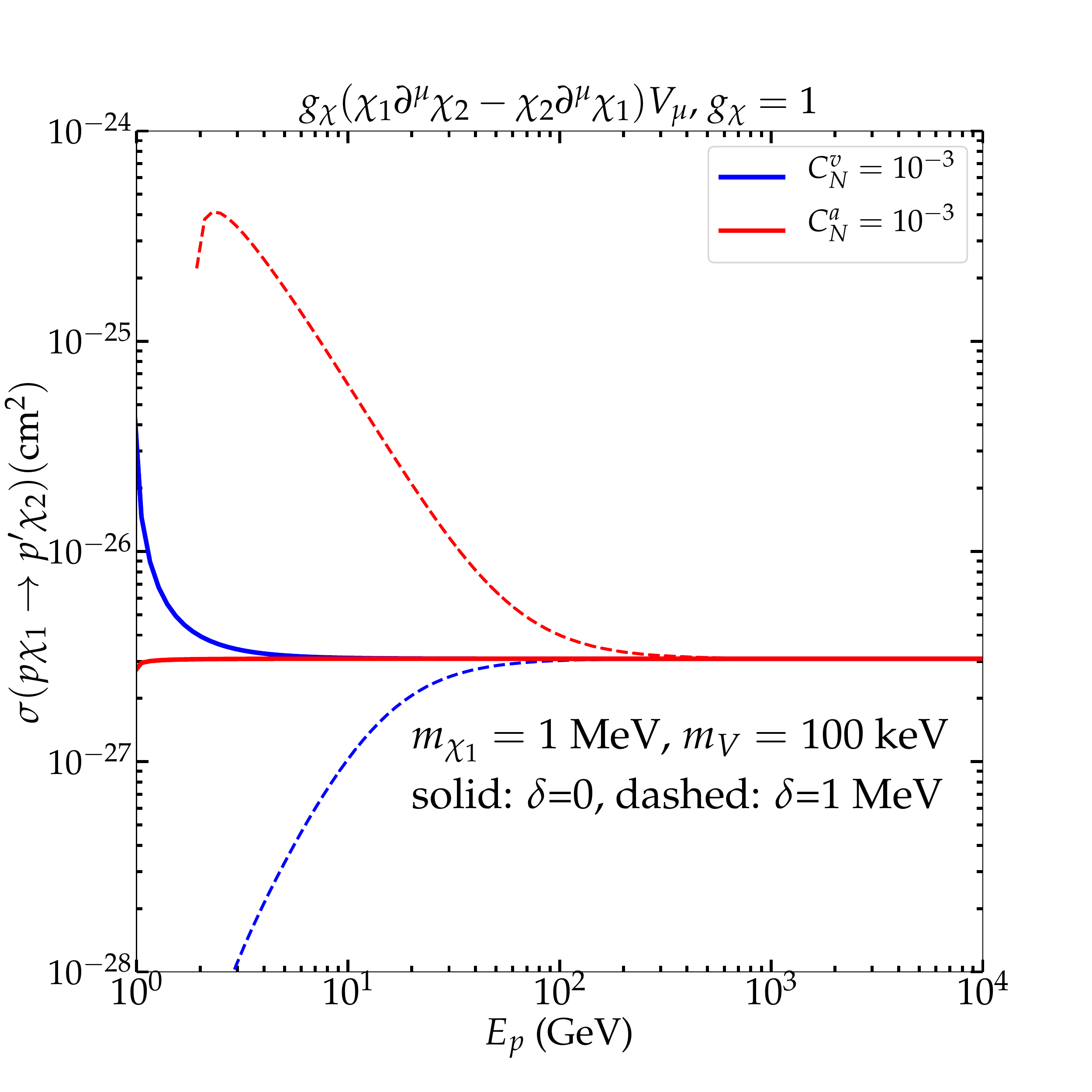}
\label{fig:Ep_Xsec_scalar_cacv_a}
}
\subfloat[$\mathcal{L}^s_{2}$]{
\includegraphics[width=0.48\textwidth]{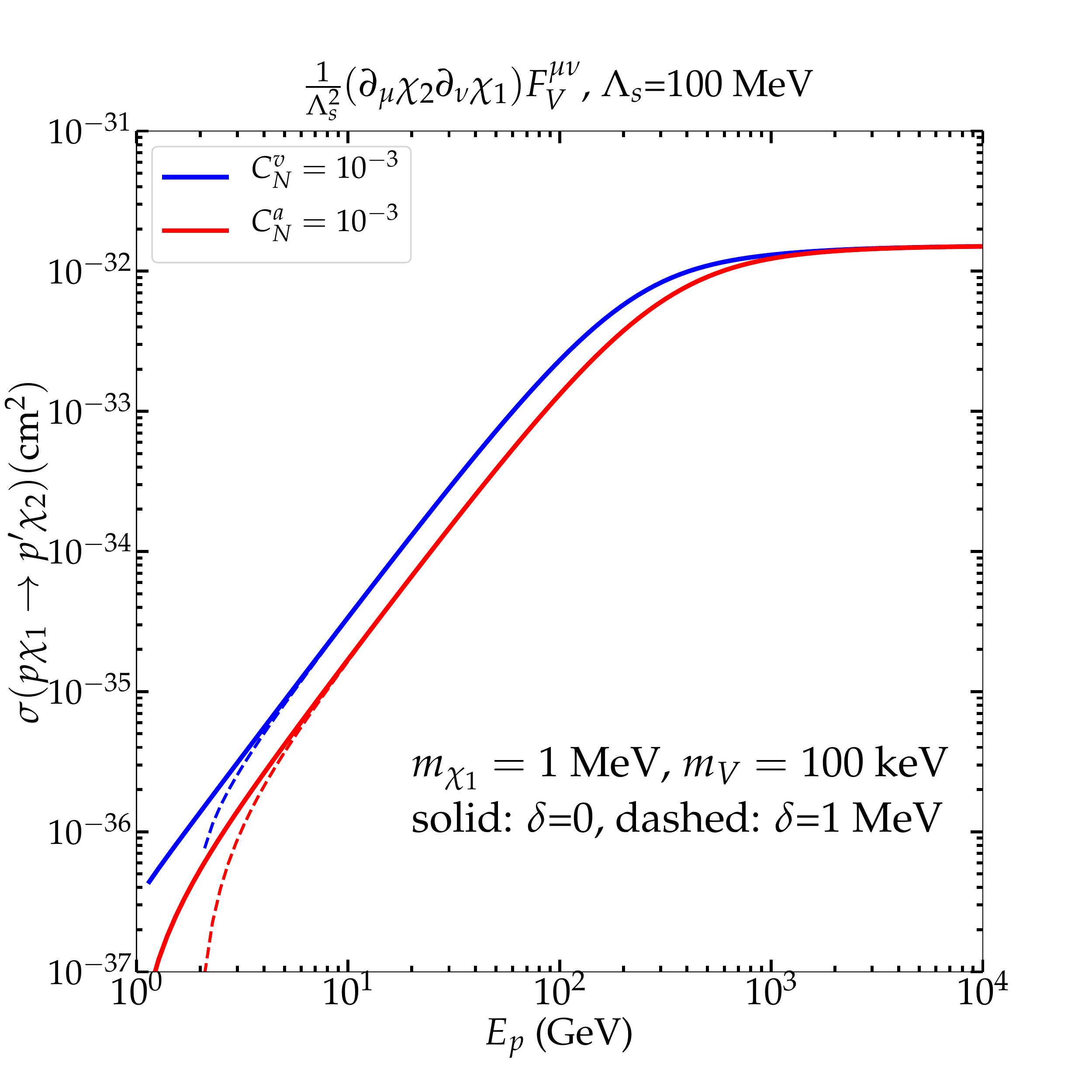}
\label{fig:Ep_Xsec_scalar_cacv_b}
}
\caption{
The integrated cross section between scalar DM and an incoming proton as a function of its energy $E_p$.
A nonzero $\delta$ contributes to an enhancement for Lagrangian $\mathcal{L}^s_{1}$ with SM axial vector interaction.
For $\mathcal{L}^s_{2}$, the resulting difference between the SM vector and axial vector interaction is tiny.
}
\label{Fig:Ep_Xsec_scalar_cacv}
\end{centering}
\end{figure*}

Regarding the interactions between the spin-zero inelastic scalar DM and the vector mediator,
we also consider two benchmarks:
\begin{eqnarray}
\mathcal{L}^s_{1} &=& g_\chi(\chi_1\partial^\mu \chi_2 - \chi_2\partial^\mu \chi_1) V_\mu,
\label{eq:s1}\\
\mathcal{L}^s_{2} &=& \frac{1}{\Lambda_s^2}(\partial_\mu \chi_2 \partial_\nu \chi_1) F_V^{\mu\nu}.
\label{eq:s2}
\end{eqnarray}
As a comparison, the dipole-like interaction Eq.~\eqref{eq:s2} is considered.
The index $s$ denotes scalar DM.
As mentioned above, only the product of the DM and SM couplings appears in the cross section.
We therefore only alter the SM coupling constant $C^{v}_N$ and $C^{a}_N$ in Eq.~\eqref{eq:f1} and fix the coupling constants of the dark sector as
$g_{\chi} = 1$ and $\Lambda_s = 100\ \rm{MeV}$.

In Fig.~\ref{Fig:Ep_Xsec_scalar_cacv}, we compare the integrated cross section for $\mathcal{L}^s_{1}$ (left panel)
and $\mathcal{L}^s_{2}$ (right panel). The color scheme is the same as Fig.~\ref{Fig:Ep_Xsec_fer_cacv}.
Fixing the vector interaction of SM part, $\mathcal{L}^s_{1}$
leads to an almost identical cross section as the fermionic case $\mathcal{L}^f_{1}$ by comparing the blue lines in Fig.~\ref{fig:Ep_Xsec_fer_cacv_a} and \ref{fig:Ep_Xsec_scalar_cacv_a}. This result is due to the similarity of their amplitude squared structure.
Conversely, for fixing the SM axial vector interaction, the $\delta=0$ case for $\mathcal{L}^s_{1}$ (red solid line) differs from the one for $\mathcal{L}^f_{1}$, where the former is flat, but an enhancement appears in the latter.
This result is due to the nontrivial behavior of amplitude squared of axial vector interaction with the unitary gauge.
When fixing the SM axial vector, we observe that both the DM interactions $\mathcal{L}^f_{1}$ and $\mathcal{L}^s_{1}$ lead to cross sections with the same order using the same coupling strength if $E_p$ increases sufficiently. Also, when $\delta=0$, the enhancement of $\mathcal{L}^s_{1}$ vanishes but remains for $\mathcal{L}^f_{1}$.
This enhancement is actually only important for a light mediator mass. For dipole form interaction $\mathcal{L}^s_{2}$, as shown in Fig.~\ref{fig:Ep_Xsec_scalar_cacv_b},
the cross section is generally suppressed at the lower $E_p$ region.
In addition, the inelastic scattering cross section does not differ markedly from
that of elastic scattering.

\begin{figure*}[htbp]
\begin{centering}
\subfloat[$\mathcal{L}^s_{1}$]{
\includegraphics[width=0.48\textwidth]{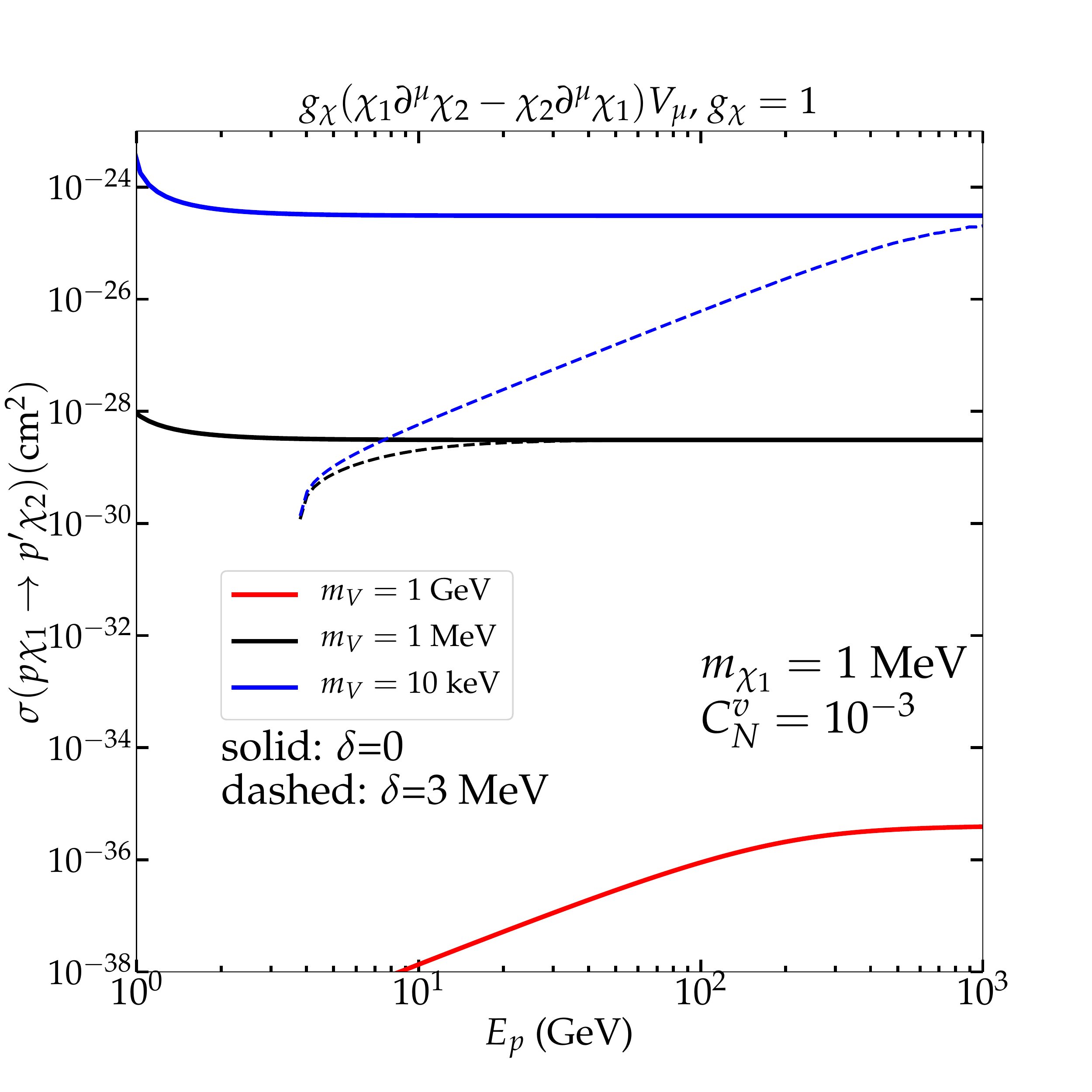}
\label{fig:Ep_Xsec_s_a}
}
\subfloat[$\mathcal{L}^s_{2}$]{
\includegraphics[width=0.48\textwidth]{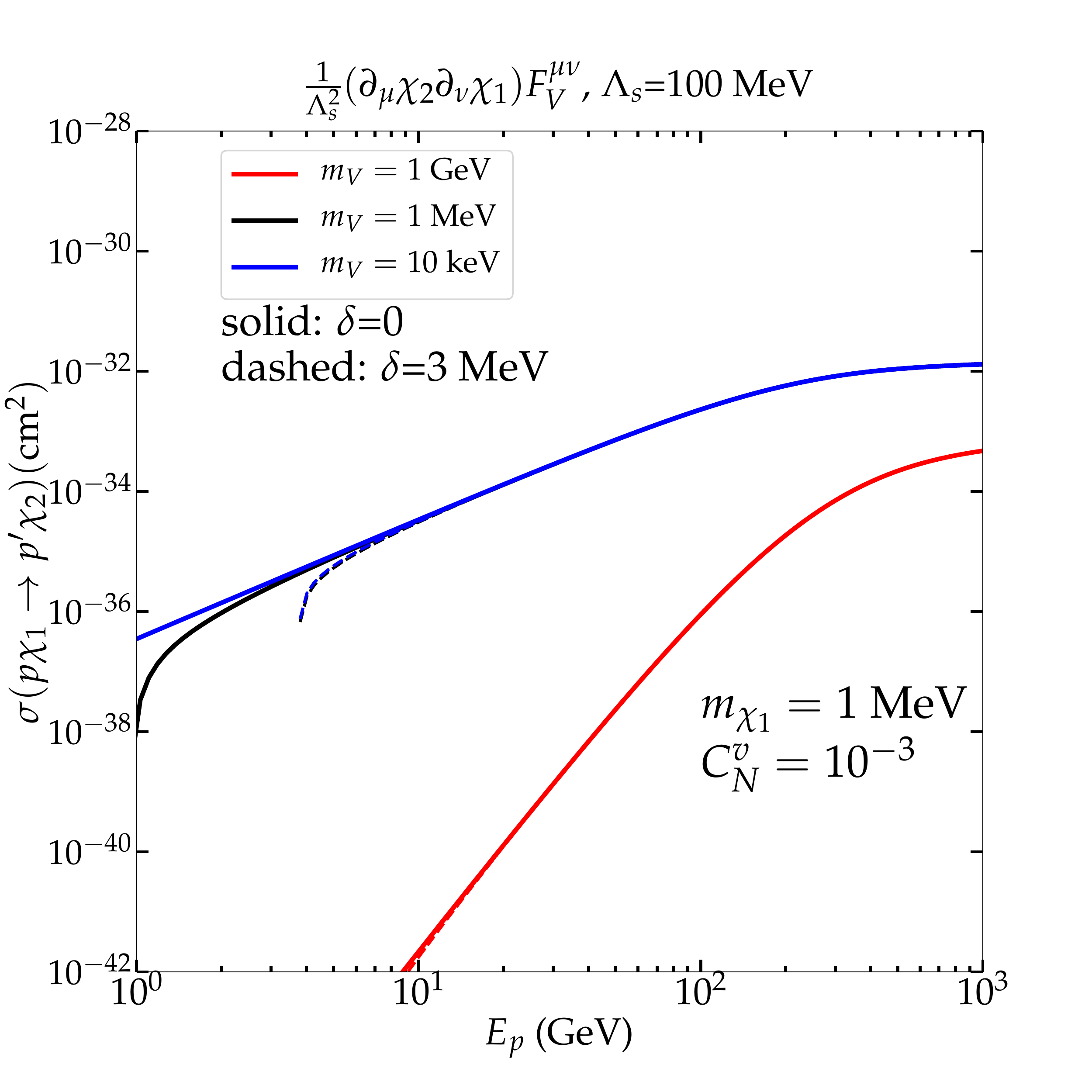}
\label{fig:Ep_Xsec_s_b}
}
\caption{
Same as in Fig.~\ref{Fig:Ep_Xsec_fer_mv} but for scalar DM scenario.
The left and right panels are for the interactions $\mathcal{L}^s_{1}$ and $\mathcal{L}^s_{2}$, respectively.
}
\label{Fig:Ep_Xsec_s}
\end{centering}
\end{figure*}

In Fig.~\ref{Fig:Ep_Xsec_s}, we plot the inelastic scalar DM scenario with the same scheme as in Fig.~\ref{Fig:Ep_Xsec_fer_mv}.
With a similar amplitude squared structure, Fig.~\ref{fig:Ep_Xsec_fer_mv_a} and \ref{fig:Ep_Xsec_s_a} are almost identical,
even if their spins are different. For $\mathcal{L}^s_{2}$ as shown in Fig.~\ref{fig:Ep_Xsec_s_b},
we can see that the cross section is not sensitive
to $\delta$ at all.
However, the cross section is sensitive to $m_V$ only for large $m_V$ (red lines), where its value is markedly suppressed.

%%%%%%%%%%%%%%%%%%%%%%%%%%%%%%%%%%%%%%%%%%%%%%%%%%%%%%%%%%%%%%%%%%%%%%%
\section{Detection of vDM and CRDM}
\label{sec:DMCR}
%%%%%%%%%%%%%%%%%%%%%%%%%%%%%%%%%%%%%%%%%%%%%%%%%%%%%%%%%%%%%%%%%%%%%%%

In this section, we first review the formulas of traditional inelastic DM scattering with the target nuclei.
vDM is present around the Earth with a local density $\rho_0=0.3\gev \cdot {\rm cm}^{-3}$,
and its velocity distribution can be simply described by a soft truncated Maxwell-Boltzmann form.
In the second part of this section, the inelastic CRDM fluxes $d\phi_\chi/dT_\chi$ during the CR-DM collisions, 
which can be altered with respect to different values of $m_{\chi_1}$ and $\delta$, are investigated.
Finally, we estimate the detected inelastic DM event rate $\mathcal{R}$
of the CRDM scattering with the target nuclei.

\subsection{Traditional detection of the virialized DM}
\label{sec:vDM}
We first consider the case of vDM. The Earth sweeps the local DM halo,
where DMs are virialized, and their velocity $v$ can be well described by
a Maxwell-Boltzmann distribution $f(v)$.
Following the convention of Ref.~\cite{Liu:2017kmx},
the differential event rate of scattering between DM and the target nucleus
\textit{per unit detector mass} with respect to the recoil energy $Q$
can be written as:
\begin{eqnarray}
\frac{{\rm d}\mathcal{R}}{{\rm d}Q} &=& \sum_{\mathcal{T}} \xi_\mathcal{T} \frac{\rho_{0}}{m_{\chi_1} m_{\mathcal{T}} }
 \int_{v > v_{\rm min}(Q)} \,  v  f(v)\frac{{\rm d}\sigma_{\chi\mathcal{T}}}{{\rm d}Q}\, d^3v,
\label{Eq:dndQ}
\end{eqnarray}
where the parameter $\xi_\mathcal{T}$ is defined as:
\begin{equation}
\xi_\mathcal{T} =  \frac{\eta_\mathcal{T} m_\mathcal{T}}{\sum\limits_{\mathcal{T}} \eta_{\mathcal{T}} m_\mathcal{T}}. 
\label{Eq:xiT}
\end{equation}
The isotope fraction $\eta_\mathcal{T}$ can also be found online\footnote{\url{https://www.webelements.com/xenon/}}.
The target mass $m_\mathcal{T}$ depends on the material used in the detectors.
The momentum transfer $\bf q$ differs from the recoil energy $Q$
but are related by ${\bf q}^2=2 m_\mathcal{T} Q$.
For a mass splitting $\delta$ between $\chi_1$ and $\chi_2$,
the kinetic phase space in the inelastic process will be restricted
by a minimum velocity, which is determined by taking the limits of small $\delta$ and $v$:
\begin{equation}
v_{\rm min}(Q)=\frac{1}{\sqrt{2 m_\mathcal{T} Q}} \left( \frac{m_\mathcal{T} Q}{\mu_{\chi N}} +\delta \right),
\label{Eq:minv}
\end{equation}
where $\mu_{\chi N}$ is the dark matter-nucleon reduced mass.
Taking the nonrelativistic limit $T_{\chi_1}=m_{\chi_1} v^2/2$ and one-dimensional velocity distribution,
Eq.~\eqref{Eq:dndQ} in terms of
the DM fluxes $d\phi_\chi/dT_\chi=\rho_0 f(v)/m_{\chi_1}^2$ returns the general form:
\begin{eqnarray}
\frac{{\rm d}\mathcal{R}}{{\rm d}Q} &=& \sum_{\mathcal{T}}  \frac{\xi_\mathcal{T} }{ m_{\mathcal{T}} }
 \int_{T_\chi[v_{\rm min}(Q)]} \,  \frac{d\phi_\chi}{dT_\chi}
 \frac{{\rm d}\sigma_{\chi\mathcal{T}}}{{\rm d}Q}\, dT_\chi.
\label{Eq:dndQ_flux}
\end{eqnarray}
In the limit $v\to 0$, the cross section $\rm{d}\sigma_{\chi\mathcal{T}}/\rm{d}Q$ includes spin-independent (SI) and spin-dependent (SD) components.
The SI component can also be coherently enhanced by
target atom number square $\mathcal{A}^2$, particularly:
\begin{equation}
\frac{{\rm d}\sigma^{\rm SI}_{\chi\mathcal{T}}}{{\rm d}Q}=
\frac{{\rm d}\sigsip^{\rm SI}}{{\rm d}Q}\times
\frac{\mu^2_{\mathcal{A}}}{\mu^2_{p}} \times \left[ \mathcal{Z} + \frac{f_n}{f_p}
\left({\mathcal{A} - \mathcal{Z}} \right) \right]^2
\times F^2(Q,\mathcal{A},\mathcal{Z})\, ,
\label{eq:sigmaXN_vDM}
\end{equation}
where $\mu^2_{p}$ and $\mu^2_\mathcal{A}$ are the DM-proton and DM-atom reduced masses, respectively, and for isospin conservation, $f_p=f_n$.
When using a velocity- and spin-independent cross section in this study,
we only take the Helm type form factor for $F^2(Q,\mathcal{A},\mathcal{Z})$
as Ref.~\cite{Lewin:1995rx}.

Finally, theoretically predicted events can be compared with experimental measurements
by including the efficiency of the underground detector $\epsilon(Q)$
to account for the experimental analysis.
Thus, the total event rate is:
\begin{equation}
\mathcal{R}=\int^{\infty}_{0}\epsilon(Q)\frac{{\rm d}\mathcal{R}}{{\rm d}Q} dQ.
\label{Eq:NRevents}
\end{equation}

\subsection{Inelastic production of CRDM}

The vDMs near the Earth are non-relativistic, and
their mean and escape velocities are approximately $240~{\rm km}\ s^{-1}$ and $540~{\rm km}\ s^{-1}$~\cite{Huang:2016,XENON:2017vdw,XENON:2018voc}.
However, DM may be accelerated by high-energy CR protons toward the Earth
with a relativistic velocity.
If the masses of $\chi_1$ and $\chi_2$ differ by a small mass splitting $\delta$, such a high energy transfer
from CR can excite $\chi_1$ to $\chi_2$.
The inelastic DM collision with the CR proton can be described as:
\begin{equation}
  p+\chi_1 \to p^\prime+\chi_2 \to p^\prime +\chi_1 + V,
  \label{eq:process}
\end{equation}
where $p$ is a cosmic proton and $V$ is a $Z_2$ even boson mediated between
SM and dark sectors.
To ensure that $V$ is produced on the shell, we require $\delta>m_V$ throughout this study.
Compared with the nonrelativistic vDM originating from our neighborhood,
the CRDM produced by $p\chi_1$ scattering may occur everywhere in our galaxy.
Similar to Ref.~\cite{Bringmann:2018cvk,Ema:2018bih},
the DM flux caused by $p\chi_1$ scattering is:
\begin{equation}
\frac{d \phi^{\rm MW}_{\chi_1}}{dT_{\chi_1}} = \int d\Omega \int_{\rm l.o.s.} d\ell \int dE_p \frac{\rho_\chi(r)}{m_{\chi_1}}\frac{d\phi_p}{dE_p}
\frac{d\sigma_{p\chi_1\rightarrow p'\chi_1 V}}{dT_{\chi_1}},
\label{eq:flx_chi}
\end{equation}
where we take the Navarro-Frenk-White halo profile for illustration
and the DM halo density is defined as $\rho_\chi$.
The integration is performed along the line of sight (l.o.s.).
The distribution $d\sigma_{p\chi_1\rightarrow p'\chi_1 V}/dT_{\chi_1}$ describes the differential cross section
of the collision process given in Eq.~\eqref{eq:process} with respect to kinetic energy of the final state $\chi_1$.
The differential flux of CR protons $d\phi_p/dE_p$ is in units of
${\rm GeV^{-1}~cm^{-2}~s^{-1}~sr^{-1}}$.

In Ref.~\cite{Guo:2020oum}, they compared two different CR proton spatial distributions:
one with a uniform and isotropic distribution in a cylinder, and
the other with the actual simulation resulting from \texttt{GALPROP}~\cite{Strong:1998pw}.
However, the differences in the accelerated DM fluxes between these two results are small.
Therefore, we can follow Refs.~\cite{Ema:2018bih,Cappiello:2019qsw,Guo:2020drq,Guo:2020oum}
to assume that the CR proton distribution is uniform and isotropic
in a cylinder with radius $R= 10$ kpc and half-height $h=1$ kpc. The spectra for protons and helium are taken from Ref.~\cite{Boschini:2017fxq} for $E_p \lesssim 10^6\gev$ (below the first knee). We can also use the CR fluxes
approximately described by a broken power law $d\phi_p/dE_p \propto E_p^{-\gamma}$, where
$\gamma\approx 3$ for $10^6 \lesssim E_p \lesssim 2\times 10^{8}\gev$ (below the second knee), and
$\gamma \approx 3.3$ for $2\times 10^8 \lesssim E_p \lesssim 3\times 10^{9}\gev$ (below the ankle).
Also, we can neglect the $E_p>3\times 10^{9}\gev$ flux due to Greisen-Zatsepin-Kuzmin
cutoff \cite{Zyla:2020zbs}.
Because we only take the spatial independent $d\phi_p/dE_p$,
we can simplify the standard DM fluxes Eq.~\eqref{eq:flx_chi}
by integrating the DM halo density:
\begin{equation}
\frac{d \phi^{\rm MW}_{\chi_1}}{dT_{\chi_1}} = \frac{\rho_0}{m_{\chi_1}}\times D_{\rm eff} \times
\sum_{i=p,{\rm He}}
\int dE_i \frac{d\phi_i}{dE_i}
G^2_i(2 m_{\chi_1} T_{\chi_1})
\frac{d\sigma_{p\chi_1\rightarrow p'\chi_1 V}}{dT_{\chi_1}},
\label{eq:x1flx}
\end{equation}
where $D_{\rm eff}$ is the effective length:
\begin{equation}
D_{\rm eff}=
\int  d\Omega \int_{\rm l.o.s.} \frac{\rho[r(\ell,\Omega)]}{\rho_0} d\ell.
\label{eq:deff}
\end{equation}
The $G^2_i(Q^2)$ here is simply taken in its dipole form:
\begin{equation}
G^2_i(Q^2)=\left[ 1+\frac{Q^2}{\Lambda_i^2}\right]^{-4},
\label{eq:xpform}
\end{equation}
where $\Lambda_p=770\mev$ and $\Lambda_{\rm He}=410\mev$.

Because the process used in Eq.~\eqref{eq:process} is a $2\to 3$ process with on-shell produced $\chi_2$, we can simply used a narrow width approximation
to break down the Feynman diagram to $p\chi_1\to p'\chi_2$ and $\chi_2 \rightarrow \chi_1 V$.
Therefore, the distribution of the cross section can be written as:
\begin{eqnarray}
\frac{d\sigma_{p\chi_1\rightarrow p'\chi_1 V}}{dT_{\chi_1}}=
\int \frac{d\sigma_{p\chi_1\rightarrow p'\chi_2}}{dT_{\chi_2}}
\frac{dT_{\chi_2}}{dT_{\chi_1}}
\frac{dB_{\chi_2 \rightarrow \chi_1 V}}{d\cos{\theta'}} d\cos{\theta'},
\label{eqn:dcro_total0}
\end{eqnarray}
where $\theta'$ is the angle between the $\chi_2$ direction in the lab frame of the $p \chi_1$ and $\chi_1$ directions for $\chi_2\to \chi_1 V$ decay in the $\chi_2$ rest frame.
The differential cross section $d\sigma_{p\chi_1 \to p' \chi_2}/dT_{\chi_2}$ and expression of $d B_{\chi_2 \rightarrow \chi_1 V}$ are shown in Appendix~\ref{app:Xsec}.

Next, we need a Jacobian $dT_{\chi_2}/dT_{\chi_1}$.
Considering the process of $\chi_2 \rightarrow \chi_1 + V$, we can express the final $T_{\chi_1}$ by initial $E_{\chi_2}$ in the lab frame:
%and $T_{\chi_1}$ can be simply obtained in terms of $T_{\chi_2}$ and $\cos{\theta'}$,
\begin{eqnarray}
T_{\chi_1}=\frac{E^*_{\chi_1}E_{\chi_2}+|{\bf p^*_{\chi_1}}|\sqrt{E^2_{\chi_2}-m_{\chi_2}^2}\cos{\theta'}}{m_{\chi_2}}-m_{\chi_1},
\label{eqn:tx1(tx2)}
\end{eqnarray}
where $E^*_{\chi_1}=(m_{\chi_1}^2+m_{\chi_2}^2-m_V^2)/(2 m_{\chi_2})$
and $|{\bf p^*_{\chi_1}}|=\sqrt{E^{*2}_{\chi_1}-m_{\chi_1}^2}$ are the energy and momentum of $\chi_1$ in the $\chi_2$ rest frame.
By inverting Eq.~\eqref{eqn:tx1(tx2)}, one can obtain the expression of $T_{\chi_2}$ as:
\begin{eqnarray}
T_{\chi_2}(T_{\chi_1},\cos\theta')=
&&\frac{m_{\chi_2}E^*_{\chi_1}(m_{\chi_1}+T_{\chi_1})
}{(E^*_{\chi_1})^2 - (|{\bf p^*_{\chi_1}}|\cos{\theta'})^2} - m_{\chi_2} \nonumber \\
&&-\frac{m_{\chi_2} |{\bf p^*_{\chi_1}}|\cos{\theta'}\sqrt{(m_{\chi_1}+T_{\chi_1})^2-(E^*_{\chi_1})^2+(|{\bf p^*_{\chi_1}}|\cos{\theta'})^2}}{(E^*_{\chi_1})^2 - (|{\bf p^*_{\chi_1}}|\cos{\theta'})^2}.
\label{eqn:tx2(tx1)}
\end{eqnarray}
We can differentiate the above equation to obtain $dT_{\chi_2}/dT_{\chi_1}$:
\begin{eqnarray}
\frac{dT_{\chi_2}}{dT_{\chi_1}}=\frac{m_{\chi_2}}{(E^*_{\chi_1})^2 - (|{\bf p^*_{\chi_1}}|\cos{\theta'})^2}\times
\left[E^*_{\chi_1}-\frac{|{\bf p^*_{\chi_1}}|\cos{\theta'}(m_{\chi_1}+T_{\chi_1})}
{\sqrt{(m_{\chi_1}+T_{\chi_1})^2-(E^*_{\chi_1})^2+(|{\bf p^*_{\chi_1}}|\cos{\theta'})^2}}\right].
\label{eqn:dtx2dtx1}
\end{eqnarray}

\begin{figure*}[htbp]
\begin{centering}
\subfloat[CRDM versus vDM . ]{
\includegraphics[width=0.49\textwidth]{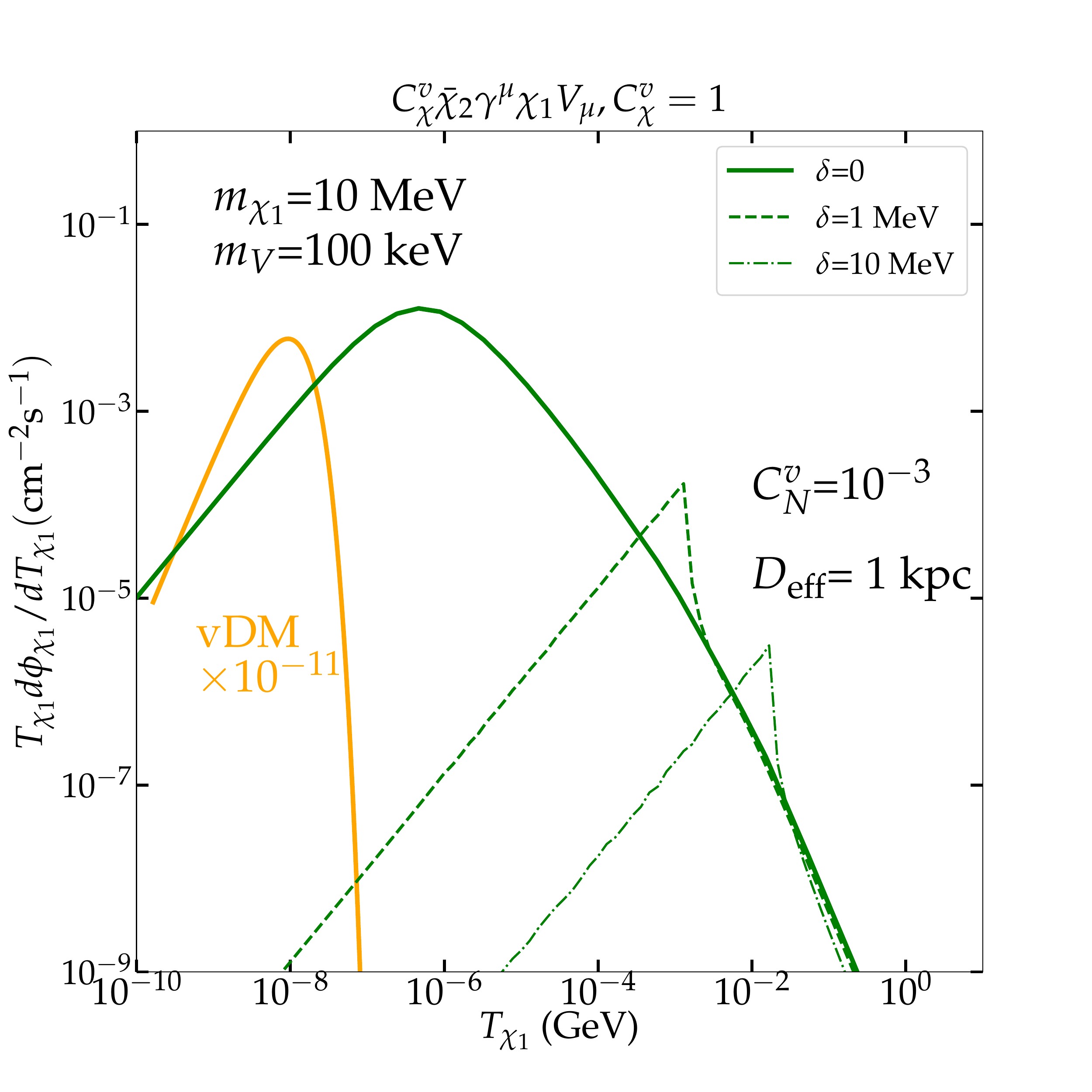}
\label{Fig:flux_fer1_a}
}
\subfloat[CRDM fluxes. ]{
\includegraphics[width=0.49\textwidth]{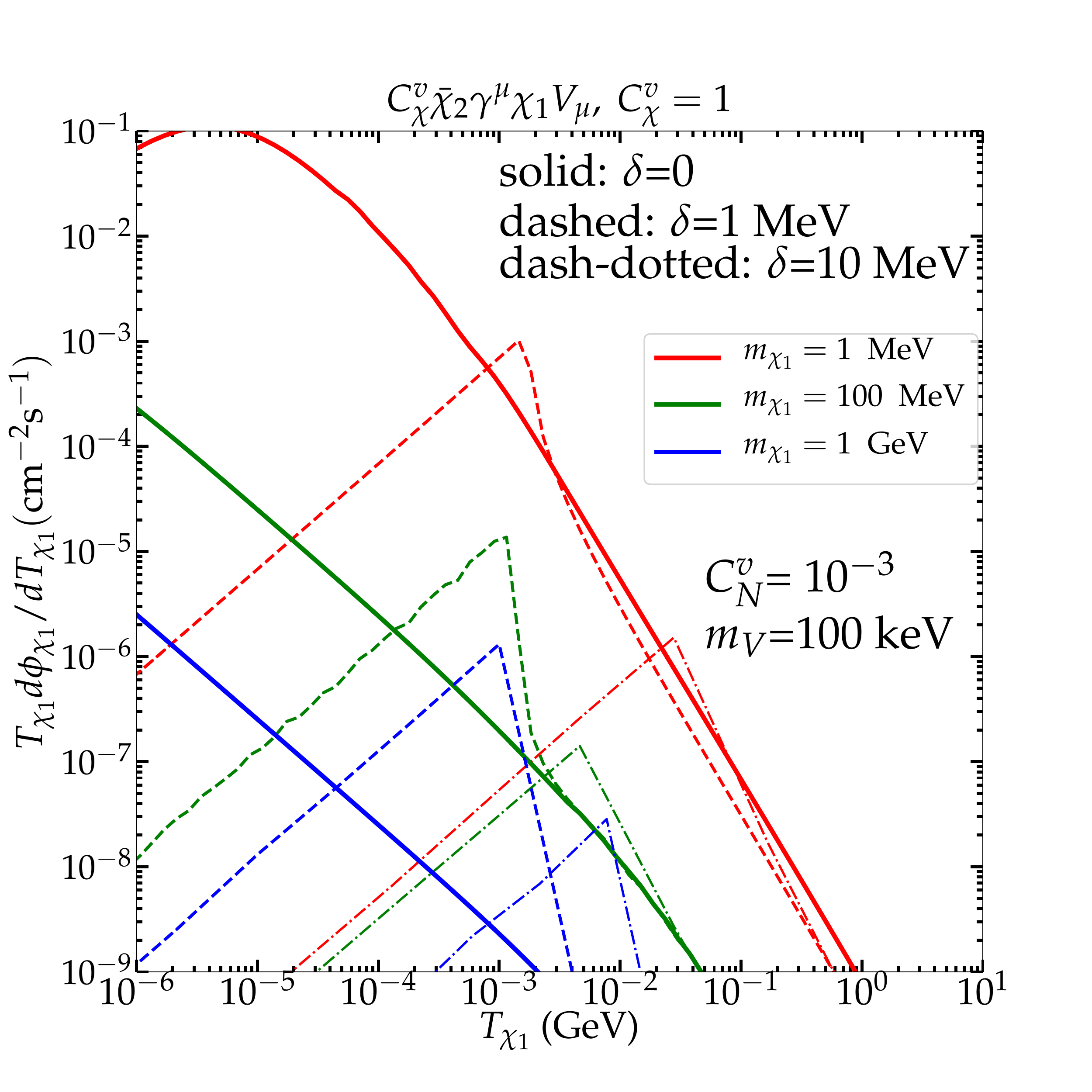}
\label{Fig:flux_fer1_b}
}
\caption{
(a) Comparison of vDM and CRDM fluxes.
Both vDM and CRDM fluxes are obtained by using $\bar{\chi_2}\gamma^\mu\chi_1 V_\mu$ interaction
with $m_{\chi_1}=10\mev$, $m_V=100\kev$, and $C_N^{a}=10^{-3}$.
The value of $D_{\rm eff}$ for CRDM is fixed to $1$~kpc as a default value.
(b) Spectra of CRDM with the VV interaction.
Three benchmarks of DM mass $1\mev$ (red lines), $100\mev$ (green lines),
and $1\gev$ (red lines) are presented.
We also plot three different mass splittings, $\delta=0$ (solid lines), $\delta=1\mev$ (dashed lines),
and $\delta=10\mev$ (dash-dotted lines).
}
\label{Fig:flux_fer1}
\end{centering}
\end{figure*}

For illustration, we calculate the fluxes of several characteristic interactions
by applying Eq.~\eqref{eqn:dcro_total0}.
Because the difference of SM axial vector and vector interaction has already been discussed (Fig.~\ref{Fig:Ep_Xsec_fer_cacv} and Fig.~\ref{Fig:Ep_Xsec_scalar_cacv}), we only display the situation of vector interaction and focus on the effects of kinematic variables. Additionally, we do not show the flux generated by $\mathcal{L}^s_1$, from which the resulting cross section is similar to the VV cross section.

In Fig.~\ref{Fig:flux_fer1}, we compare the vDM flux (orange solid line) with
CRDM fluxes (green lines) using $m_{\chi_1}=10\mev$, $m_V=100\kev$, and $C_N^{v}=10^{-3}$ (left).
We use VV interaction for CRDM as a demonstration,
but the shapes for other interactions do not differ much except for their scales.
For CRDM, we take $D_{\rm eff}=1$~kpc as a default value.
The green solid line represents elastic scattering with $m_{\chi_1}=m_{\chi_2}$, but
the dashed and dashed-dotted lines describe the mass splitting $\delta=1\mev$ and $\delta=10\mev$, respectively.
We note that a larger mass splitting makes a stronger suppression for the lower $T_{\chi_1}$.
We can clearly see that the fluxes of vDM in the Fig.~\ref{Fig:flux_fer1_a} peaks
at approximately $10^{-8}\gev$, where DM has a nonrelativistic velocity around $10^{-3}c$.
As an expected feature from CRDM, its $T_{\chi_1}$ can be comparable with $m_{\chi_1}$,
where DM is relativistic particle.
In fact, light vDM, such as $m_{\chi_1}=10\ \mev$, is not detectable for the present underground experiments due to its low kinetic energy. However, both elastic and inelastic CRDM obtain enough energy to be observed.
Also, we can see that the magnitude of fluxes of vDM can be approximately $11$ orders higher than
CRDM, which occurs because the integrated cross section of CR collision with DM is small ($\sim 10^{-27}$~cm$^{2}$),
even if the CRDM fluxes are accumulated over the line of sight, particularly $D_{\rm eff}$. 
The current powerful PandaX-4T and XENON1T detectors are thus capable of testing such events.

We also show the CRDM fluxes with three different DM masses and splittings in Fig.~\ref{Fig:flux_fer1_b}.
We use three benchmarks of DM mass $1\mev$ (red lines), $100\mev$ (green lines),
and $1\gev$ (blue lines). For the three mass splittings, the solid, dashed, and dash-dotted lines
correspond to the mass-degenerated case, $\delta=1\mev$, and $\delta=10\mev$, respectively.
We can see that all the mass-degenerated cases have smooth curves, but
those non-degenerated curves show sharp peaks at $T_{\rm sp}$.
The spectrum follows the CR spectrum at the $T_\chi>T_{\rm sp}$ region
but also has a markedly different shape from the CR shape in the $T_\chi<T_{\rm sp}$ region.
As shown in Eq.~\eqref{eq:x1flx}, the CRDM energy spectrum is the product of the distribution
$d\sigma_{p\chi_1}/dT_{\chi_1}$ and CR energy spectrum.
We also show in Figs.~\ref{Fig:Ep_Xsec_fer_cacv}, \ref{Fig:Ep_Xsec_fer_mv}, \ref{Fig:Ep_Xsec_scalar_cacv},
and \ref{Fig:Ep_Xsec_s} that the cross section is flat at high energy, but
the $\delta$ contribution dominates $d\sigma_{p\chi_1}/dT_{\chi_1}$ in the lower energy region. Thus, in the small energy region $T_\chi<T_{\rm sp}$, the spectrum is strongly affected by $d\sigma_{p\chi_1\to p'\chi_1 V}/dT_{\chi_1}$.

For the CRDM with $\delta=0$, its energy is only provided by $p \chi_1$ elastic collision.
Moreover, a final state $\chi_1$ can be boosted by $\delta$ owing to the $\chi_2$ decay after the inelastic $p \chi_1 \to p' \chi_2$ process, as also mentioned in Ref.~\cite{Bell:2021xff}.
Therefore, the CRDM flux with a larger $\delta$ can depart from
the mass-degenerate scenario toward higher energy.
\textit{Thus, $\delta$ can cause a sharp peak $T_{\rm sp}$ in the spectrum toward higher energy
but with a decline of total flux.}

\begin{figure*}[htbp]
\begin{centering}
\subfloat[Fermionic DM with $m_V=100\kev$.]{
\includegraphics[width=0.49\textwidth]{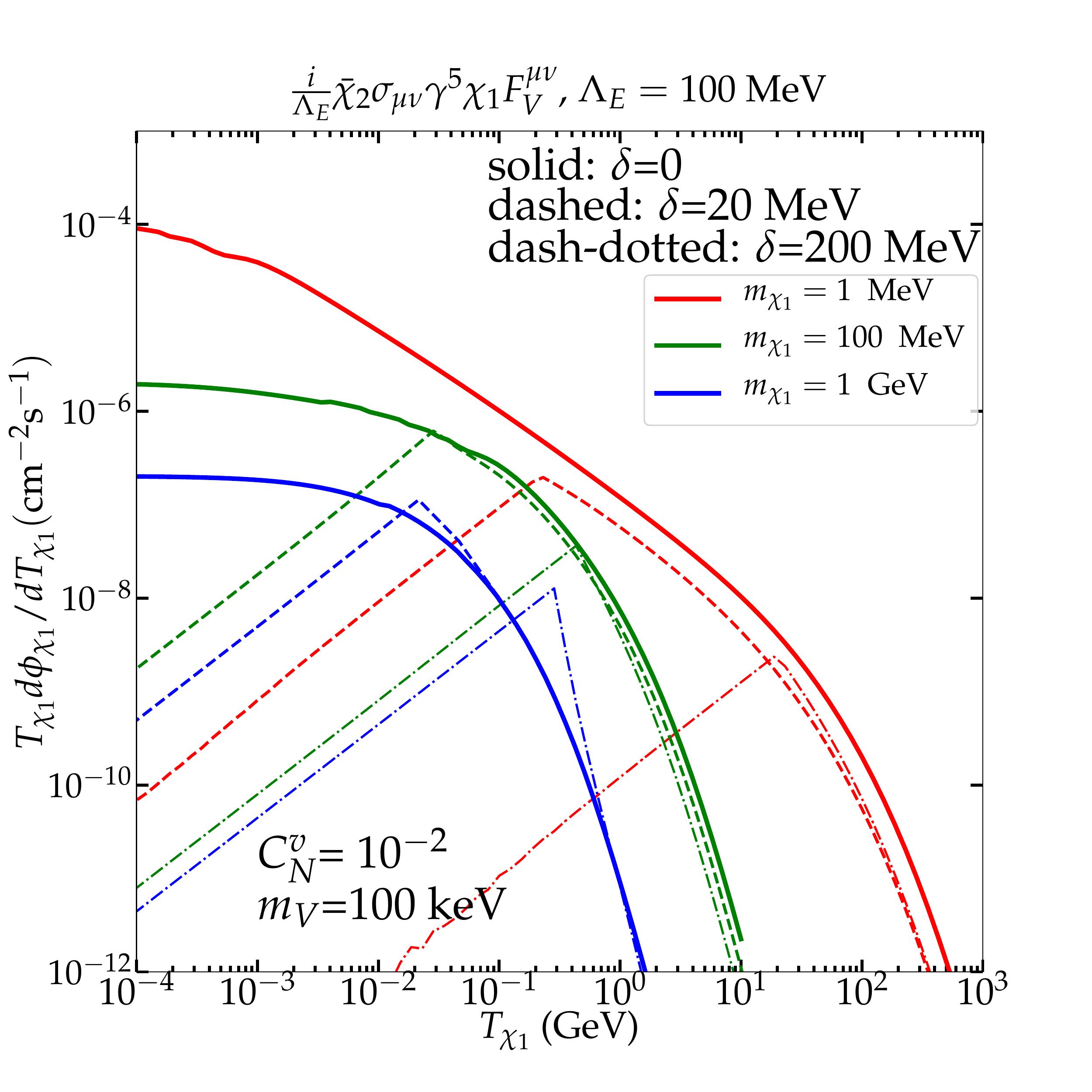}
\label{Fig:flux_f2_a}
}
\subfloat[Fermionic DM with $m_V=10\mev$]{
\includegraphics[width=0.49\textwidth]{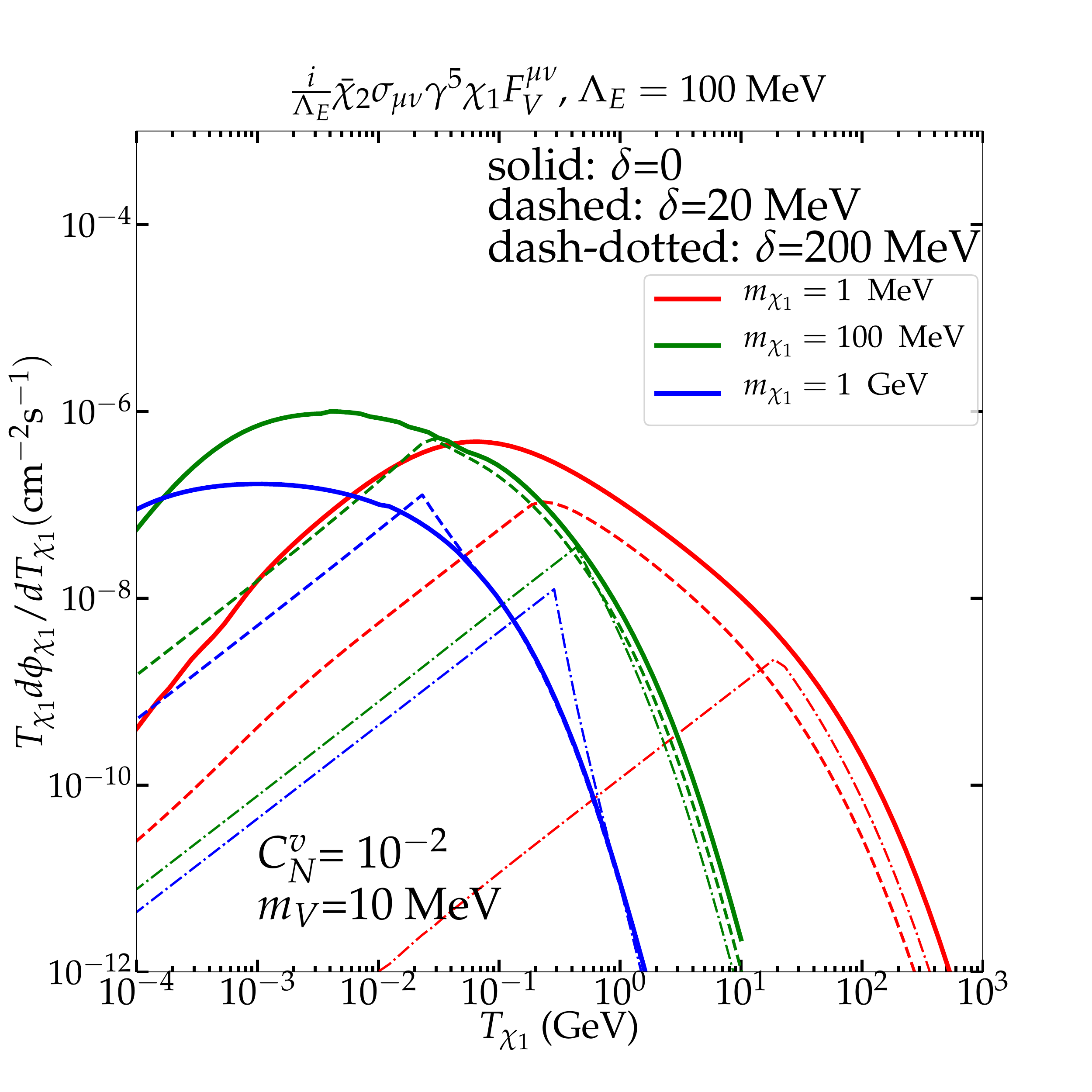}
\label{Fig:flux_f2_b}
}\\
\subfloat[Scalar DM with $m_V=100\kev$.]{
\includegraphics[width=0.49\textwidth]{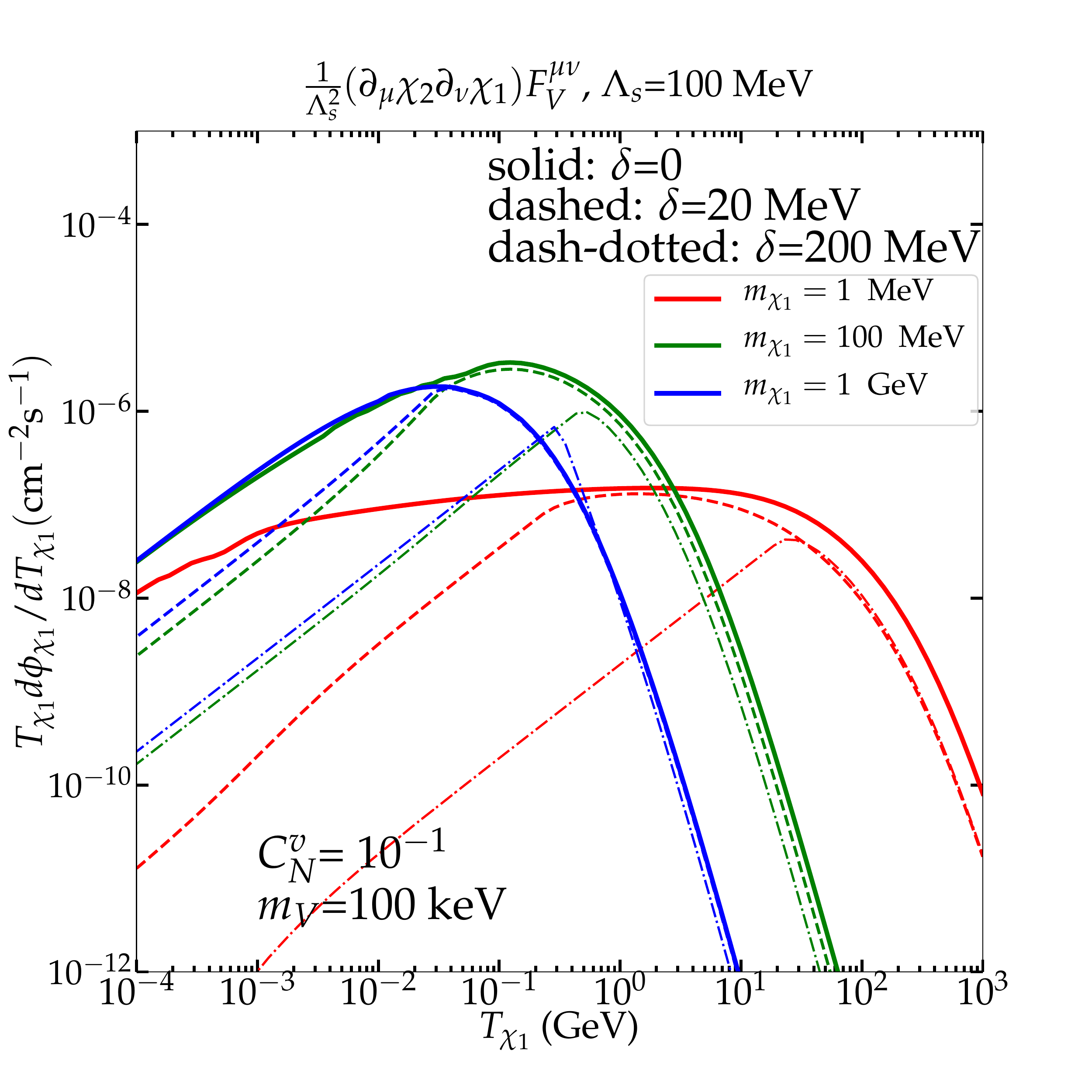}
\label{Fig:flux_s2_a}
}
\subfloat[Scalar DM with $m_V=10\mev$]{
\includegraphics[width=0.49\textwidth]{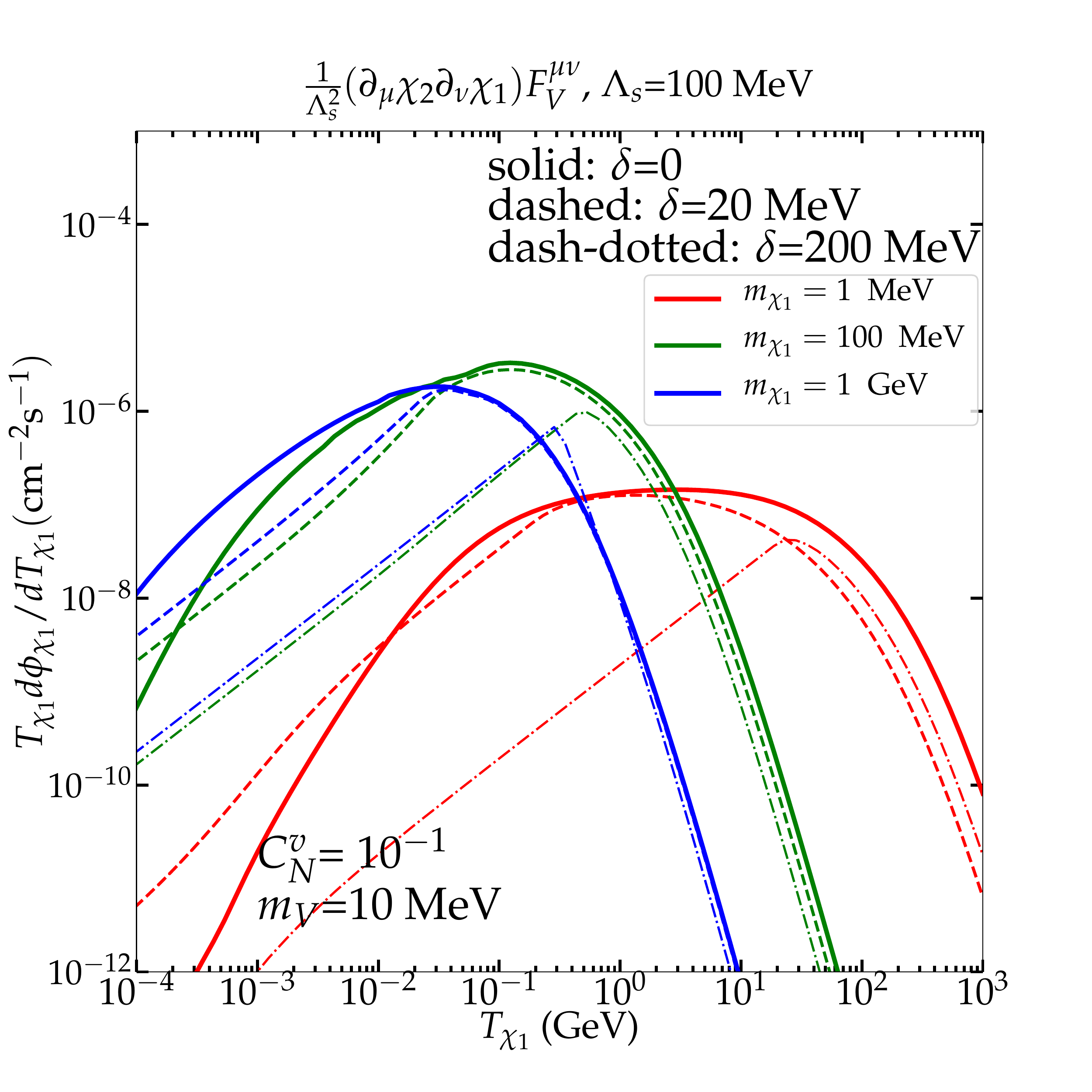}
\label{Fig:flux_s2_b}
}
\caption{Energy spectra of CRDM for $\mathcal{L}^f_{2}$ (upper panels) and $\mathcal{L}^s_{2}$ (lower panels).
The mediator masses applied here are $m_V=100\kev$ (left figures) and $m_V=10\mev$ (right figures).
The color scheme is the same as Fig.~\ref{Fig:flux_fer1_b}.
}
\label{Fig:flux_dipole}
\end{centering}
\end{figure*}

For dipole-like operators, we show the CRDM fluxes for fermionic DM cases (two upper panels) and scalar DM cases (two lower panels)
In Fig.~\ref{Fig:flux_dipole}. The left panels show the light mediator scenarios ($m_V=100\kev$), while
the right panels show the heavy mediator scenario ($m_V=10\mev$).
We use the same color scheme as Fig.~\ref{Fig:flux_fer1_b}, and the important features such as the shifts of the shark peaks
are also mentioned previously.
Comparing with the VV interaction in Fig.~\ref{Fig:flux_fer1_b},
the spectra of the elastic scattering of all dipole-like interactions are softer.
Therefore, the fluxes of the sharp peaks created by $\chi_2$ decay ($\delta \neq 0$) are generally lower
than $\delta=0$ case in the dipole-like interactions.
Finally, the flux of the scalar dipole-like DM interaction is lower than that of the fermionic interaction when using the same values of couplings. Note that $C_N^v=10^{-1}$ for the former and
$C_N^v=10^{-2}$ for the latter.

\subsection{CRDM detection rate}
\label{sec:CRDMdet}
Following the conventions of Ref.~\cite{Bringmann:2018cvk,Bondarenko:2019vrb},
we can write down the differential recoil rate per target nucleus of relativistic DM in underground detectors:
\begin{equation}
\frac{d\mathcal{R}}{d Q}=\sum_{\mathcal{T}}  \frac{\xi_\mathcal{T} }{ m_{\mathcal{T}} } \int^\infty_{T_{\texttt{min} }} dT_{\chi_1}
\frac{d\sigma_{\chi_1 \mathcal{T}}}{d Q}
\frac{d\Phi^{\rm MW}_{\chi_1}}{dT_{\chi_1}},
\label{eq:R_rel}
\end{equation}
where $T_{\texttt{min} }$ is the minimum kinetic energy of incoming DM, and
only $\mathcal{T}=$xenon is used in this study.
For those DM that strongly interact with nuclei,
$T_{\texttt{min} }$ can be varied with respect to the DM length of propagation
in the Earth and the DM-nuclei cross section in the attenuation process.
When the attenuation effect is negligible,
the $T_{\texttt{min} }$ only depends on the kinematics, as shown in Appendix.~\ref{app:kinematics}.

\begin{figure*}[htbp]
\begin{centering}
\subfloat[VV and AA interactions. ]{
\includegraphics[width=0.49\textwidth]{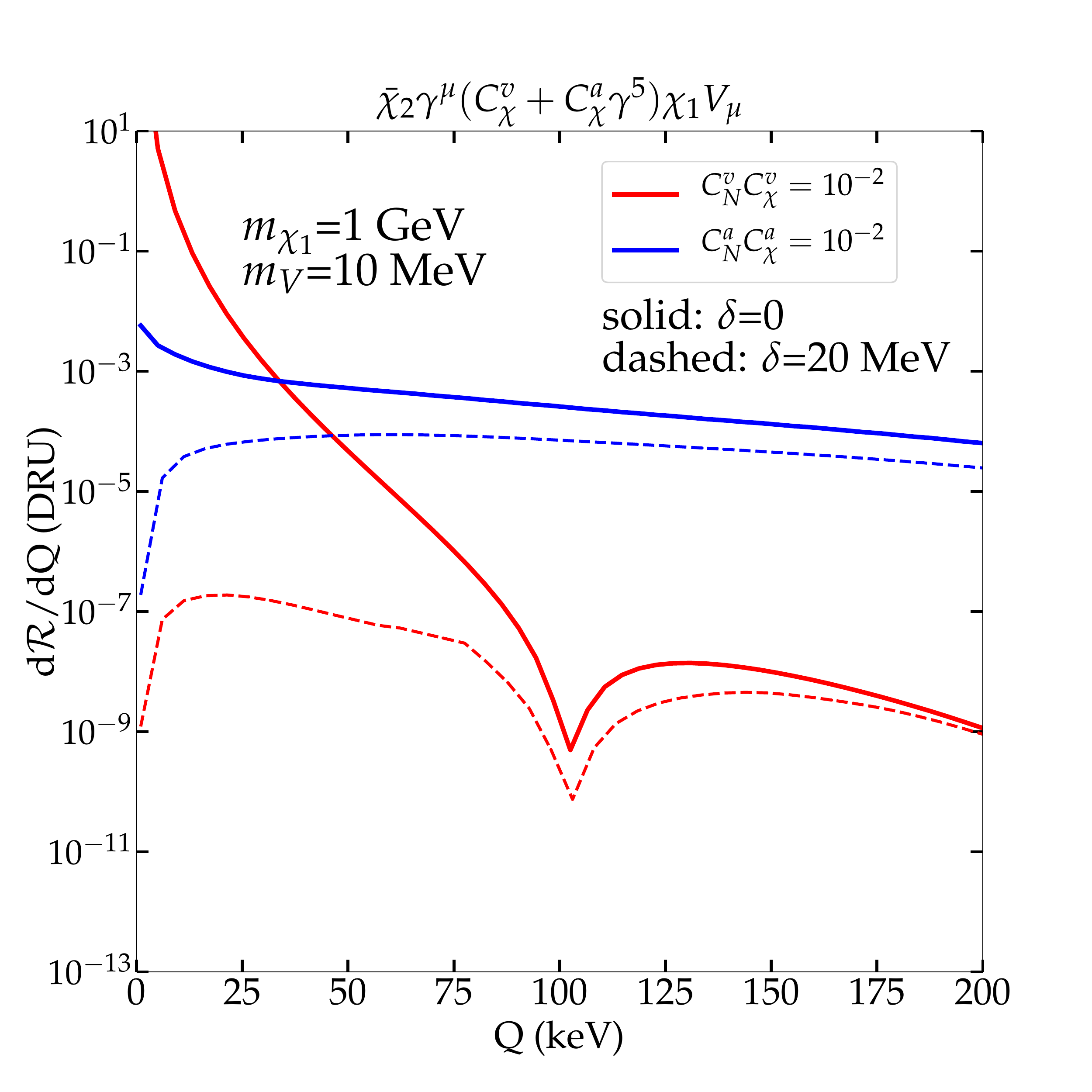}
\label{fig:dRdQ_AAVV}
}
\subfloat[ED and MD interactions. ]{
\includegraphics[width=0.49\textwidth]{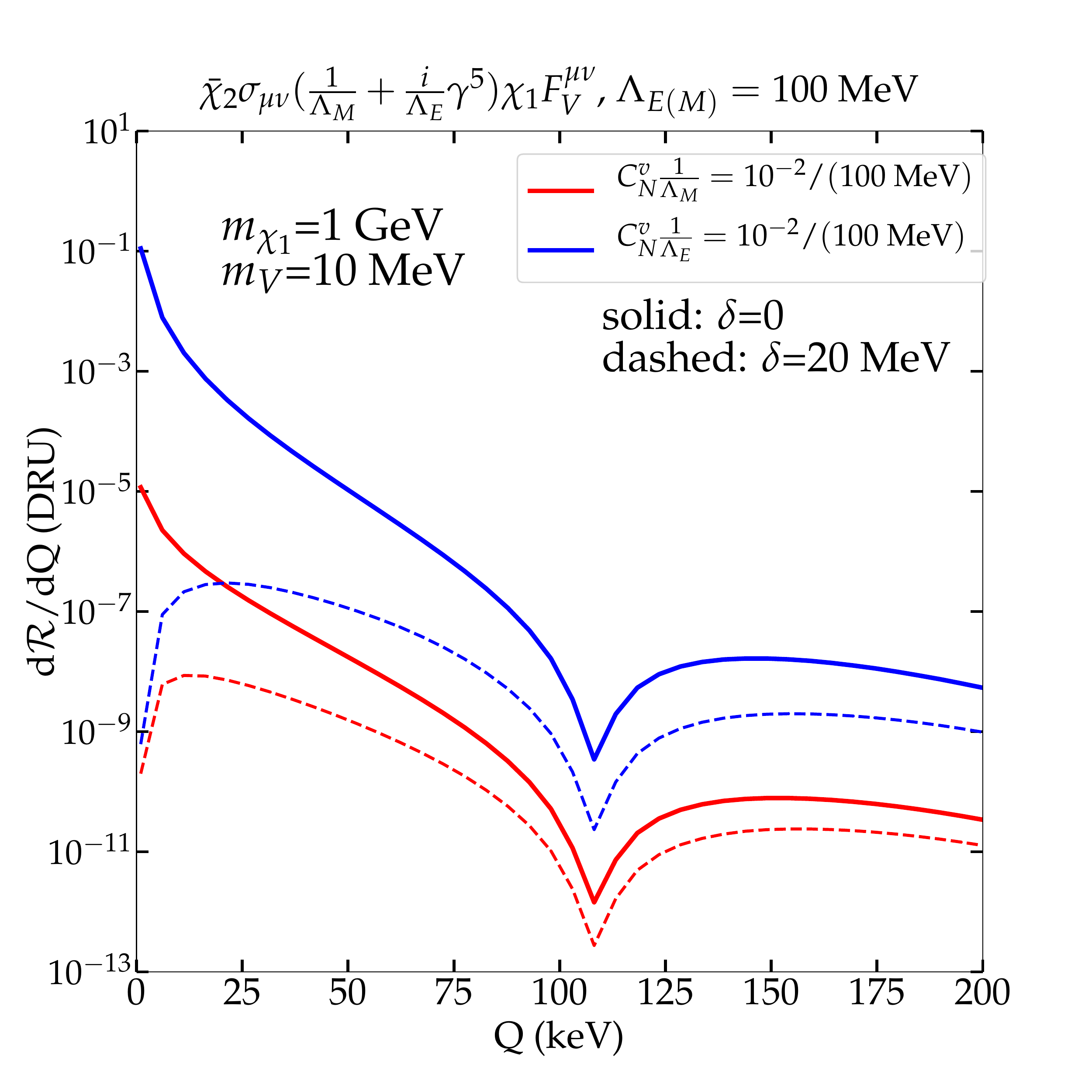}
\label{fig:dRdQ_EDMD}
}
\caption{
Detection rate $d\mathcal{R}/dQ$ (DRU or $\rm{keV}^{-1}\rm{kg}^{-1}\rm{day}^{-1}$) of fermionic CRDM $\chi-$xenon scattering for $\mathcal{L}^f_{1}$ (left)
and $\mathcal{L}^f_{2}$ (right).
The mass degenerated cases $\delta=0$ are represented as solid lines, while the
dashed lines are for $\delta=20\mev$.
}
\label{Fig:rate_fer1}
\end{centering}
\end{figure*}

As shown in Sec.~\ref{sec:vDM}, the final $d\sigma_{\chi N}/d Q$ has to be the product of the $\chi p$ cross section and
the form factor as performing in Eq.~\eqref{eq:sigmaXN_vDM}.
However, the form factor for the SI or SD cross section is derived by velocity-independent technology.
As shown in Ref.~\cite{Fitzpatrick:2012ix,Fitzpatrick:2012ib,Anand:2013yka,Bishara:2016hek,Bishara:2017nnn},
a velocity-dependent collision (e.g., dipole interaction) sometimes contains more than these two contributions.
Therefore, we use the effective operator method
developed by Ref.~\cite{Fitzpatrick:2012ix,Fitzpatrick:2012ib,Anand:2013yka,Bishara:2016hek,Bishara:2017nnn}
to extract the complete form factor for our inelastic interactions. In these studies, the leading order of the velocity contribution was used.
Considering that the form factor relates $\sigma_{\chi p}$ to $\sigma_{\chi \mathcal{T}}$, the inelastic cross section of DM-target scattering is:
\begin{eqnarray}
\frac{d\sigma_{\chi \mathcal{T}}}{d Q} (m_{\chi_1}, \delta, m_V, Q)=
\left.
\frac{d\sigma_{\chi \mathcal{T}}}{d Q}\right|_{\rm EFT}
\times \frac{\mathcal{M}_{\chi p}^2(m_{\chi_1}, \delta, m_V, Q)}{\mathcal{M}^2_{\chi p, \rm EFT}}
\label{eq:eff_formfac}
\end{eqnarray}
where the effective operator cross section $\frac{d\sigma_{\chi \mathcal{T}}}{d Q}|_{\rm EFT}$ is obtained with
the publicly available numerical code \texttt{LikeDM-DD}~\cite{Liu:2017kmx}.
$\mathcal{M}_{\chi p}^2(m_{\chi_1}, \delta, m_V)$ are calculated in Appendix~\ref{app:Xsec},
but $\mathcal{M}^2_{\chi p, \rm EFT}$ is used with $\delta=0$ and the limit $m_V\gg Q$.

Because the form factors are independent of $m_V$ and $\delta$, Eq.~\eqref{eq:eff_formfac} is valid.
The VV, AA, MD, and ED interactions in the effective theory limit correspond to the effective operators
$\mathcal{Q}^{(6)}_1$, $\mathcal{Q}^{(6)}_4$, $\mathcal{Q}^{(5)}_1$, and $\mathcal{Q}^{(5)}_2$ in Ref.~\cite{Liu:2017kmx}, respectively.
Only the fermionic DM cases are presented here because
the result for scalar interaction $\mathcal{L}^s_{1}$ is similar to that of the VV interaction, 
and $\mathcal{L}^s_{2}$ is markedly suppressed by $\Lambda_s^2$.

We now show the predicted event rate with fermionic VV (red lines in Fig.~\ref{fig:dRdQ_AAVV}),
AA (blue lines in Fig.~\ref{fig:dRdQ_AAVV}),
MD (red lines in Fig.~\ref{fig:dRdQ_EDMD}) and ED (blue lines in Fig.~\ref{fig:dRdQ_EDMD}) interactions.
We can clearly see a dip in the curves of VV, ED, and MD whose form factor contains a significant contribution
from the SI component.
The AA interaction may contain the SD component without $\mathcal{A}^2$ enhancement, but the CRDM fluxes of AA interaction are generally higher than those of VV
so that the AA event rate is not much lower than VV at the lower recoil range in the elastic case.
When $Q>40\kev$, the spectrum of AA can be even higher than that of VV by using the same parameters.
Generally, in the elastic case, the AA interaction predicts the highest event rate at high $Q$,
but VV predicts the highest at the low $Q$.
Due to the new physics scale $\Lambda_{E(M)}$ suppression,
the dipole-like interactions ED and MD generally lead to a lower rate.
Also, splitting $\delta$ can reduce the detected rate, and AA predicts the highest rate for $\delta=20\mev$.
Thus, unlike elastic scattering scenario,
the target xenon is more sensitive to detect the AA inelastic scattering cross section
than the VV interaction, even if the AA interaction may not depend on the SI form factor enhanced by coherence.

{\it \underline{Comments about the attenuation of the DM flux during propagation}:} In this study, we ignore the attenuation effect from the Earth
but focus on the exclusion limits, as in Ref.~\cite{Bell:2021xff}.
In principle, a simple version of transport equations should at least include the propagation of $\chi_1$ and $\chi_2$.
For the elastic CRDM scenario, attenuation is important
for $\sigsip>\mathcal{O}(10^{-28})$~cm$^2$~\cite{Bringmann:2018cvk,Ema:2018bih,Cappiello:2018hsu,Cappiello:2019qsw,Guo:2020drq}.
However, it is difficult to know for inelastic scattering without a simulation.
Considering the same cross section for both $\delta=0$ and $\delta>0$ scenarios,
the inelastic DM can lose its energy much more efficiently than elastic DM because of $\chi_1\to \chi_2$ excitation and $\chi_2$ decay.
Conversely, if using the same coupling for these two scenarios,
the cross section for the $\delta>0$ case may be lower than that for $\delta=0$, such as for the VV case.

Quantitatively,
a numerical code that simulates the energy distribution of $\chi_1$ and $\chi_2$
after their propagation should be developed but is beyond the scope of this study.
Our research team plans to return to this issue in the future with a novelly designed numerical code.

\section{Current constraints from PandaX-4T}
\label{sec:pandax4t}

The current, most stringent limitation of the DM-proton scattering cross section is from PandaX-4T~\cite{PandaX:2021osp}.
The DM-proton elastic scattering cross section $\sigsip$ above $3.3\times 10^{-47}$~cm$^2$ at DM mass $\sim 30\gev$ 
is excluded in $90\%$C.L. with 3.7 tons of liquid xenon target and an exposure of $0.63$ tonne$\times$year.
However, such a limit is only applicable to elastic vDM.
To apply to inelastic and relativistic DM scenarios, we must restore the limit of event rate $\mathcal{R}$
rather than that of $\sigsip$.

Thus, we can recast the PandaX-4T $90\%$ C.L. of event rate in the following.
First, we insert the published $\sigsip$ values of PandaX-4T $90\%$ C.L. into Eq.~\eqref{Eq:dndQ} and Eq.~\eqref{Eq:NRevents},
Then, the event rates for a fixed DM mass can be determined.
Actually, the published PandaX-4T exclusion plot is based on the recoil energy spectra,
which is related to the incoming DM kinetic energy.
In this study, we use the efficiency curve from Ref.~\cite{PandaX:talks},
and the window of the maximum efficiency is located for the detected recoil energies between $20\kev$ and $100\kev$.
The total event rate with an incoming DM momentum larger than threshold energy $ 30\mev$ ($m_{\chi_1}=30\gev$
with DM velocity $10^{-3} c$) is found to be nearly constant at $\mathcal{R}\sim 4/0.63$/tonne/year.
However, $\mathcal{R}$ varies rapidly when the incoming DM momentum is smaller than $30\mev$.
Thus, we use $\mathcal{R}\sim 4/0.63$/tonne/year projected onto the ($\delta$, $C_N$)
and ($m_{\chi_1}$, $\delta$) planes to show the detection capability of PandaX-4T.

\begin{figure*}[htbp]
\begin{centering}
\subfloat[$m_{\chi_1}=5\mev$.]{
\includegraphics[width=0.49\textwidth]{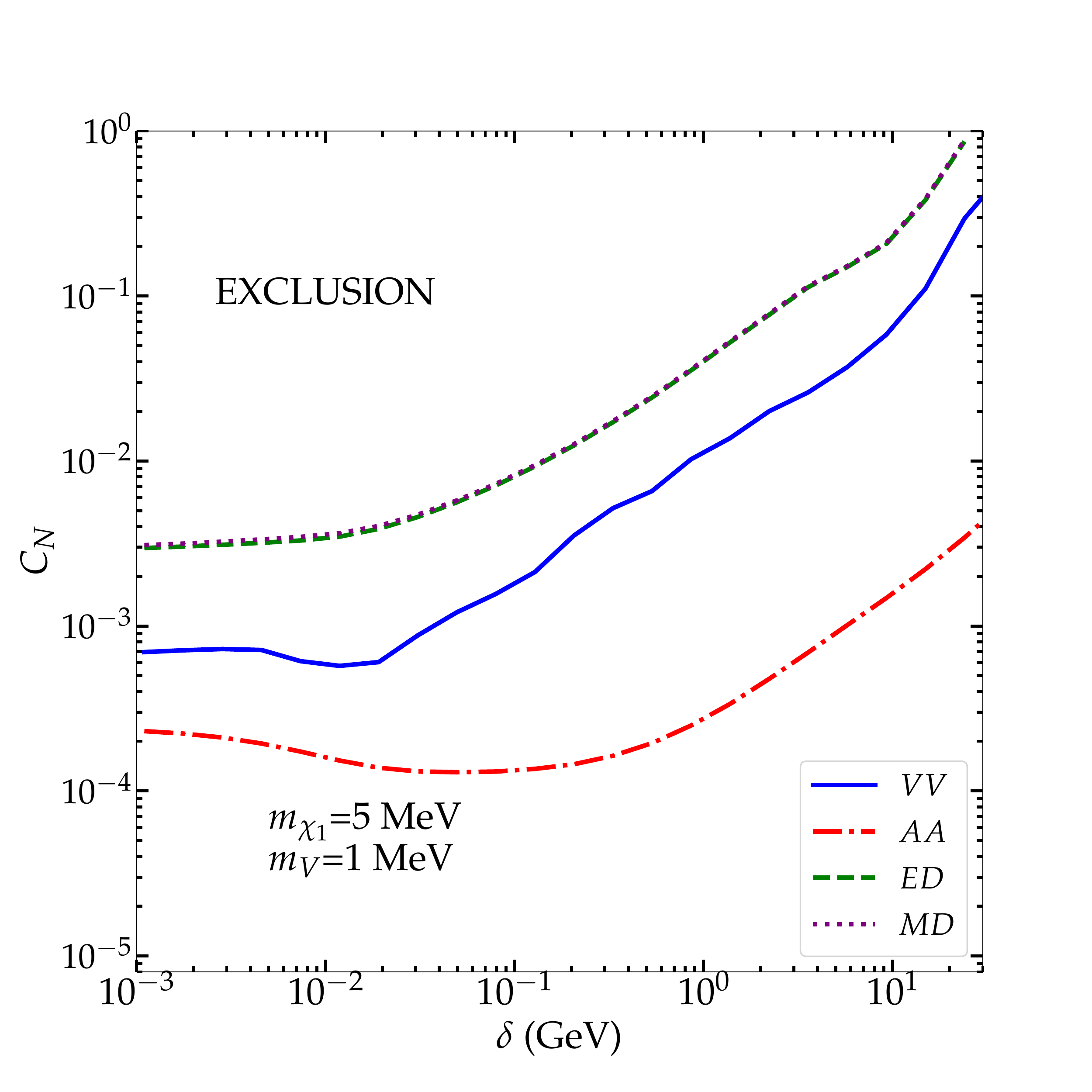}
\label{fig:delta_CN_a}
}
\subfloat[$m_{\chi_1}=1\gev$.]{
\includegraphics[width=0.49\textwidth]{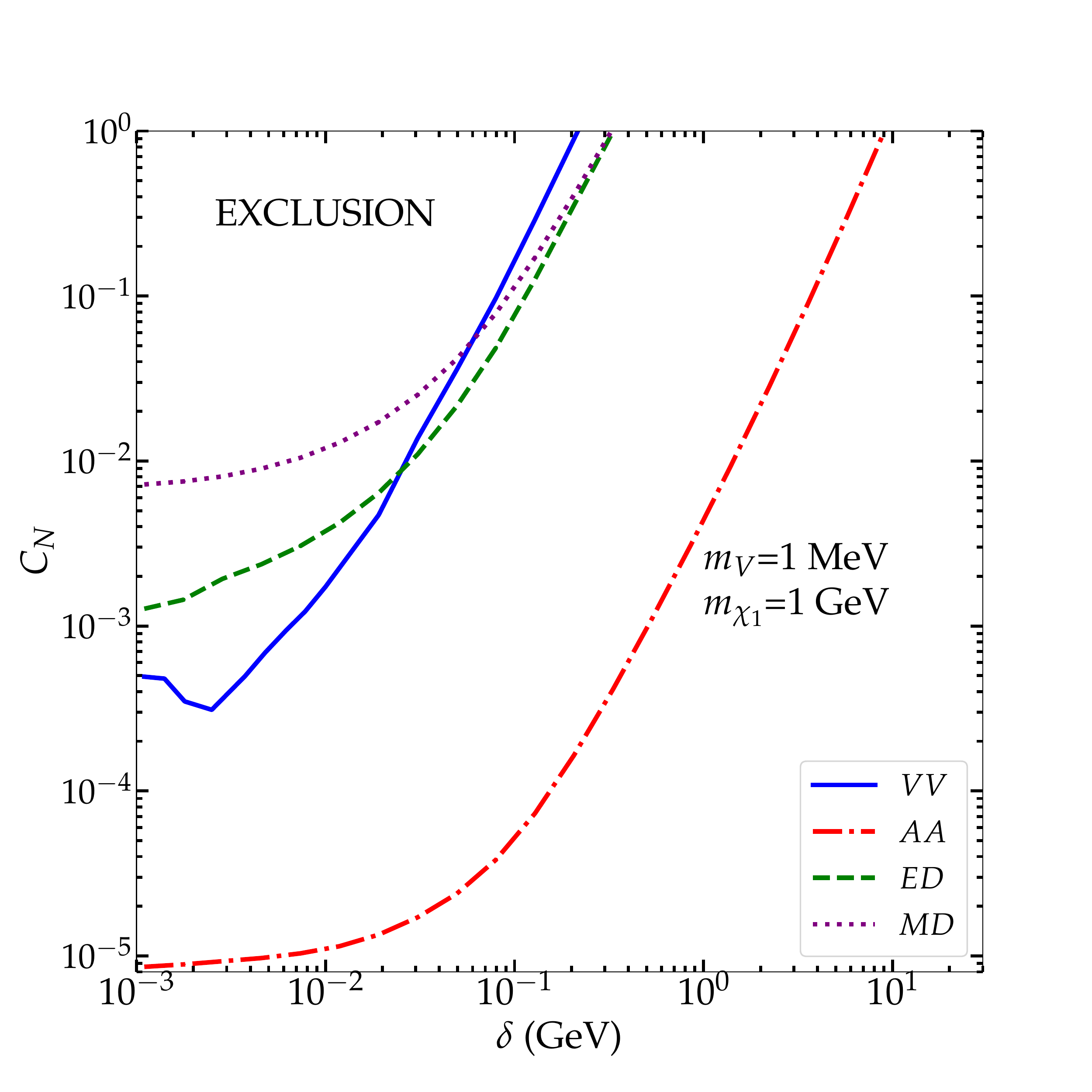}
\label{fig:delta_CN_b}
}\\
\subfloat[$m_{V}=1\mev$.]{
\includegraphics[width=0.49\textwidth]{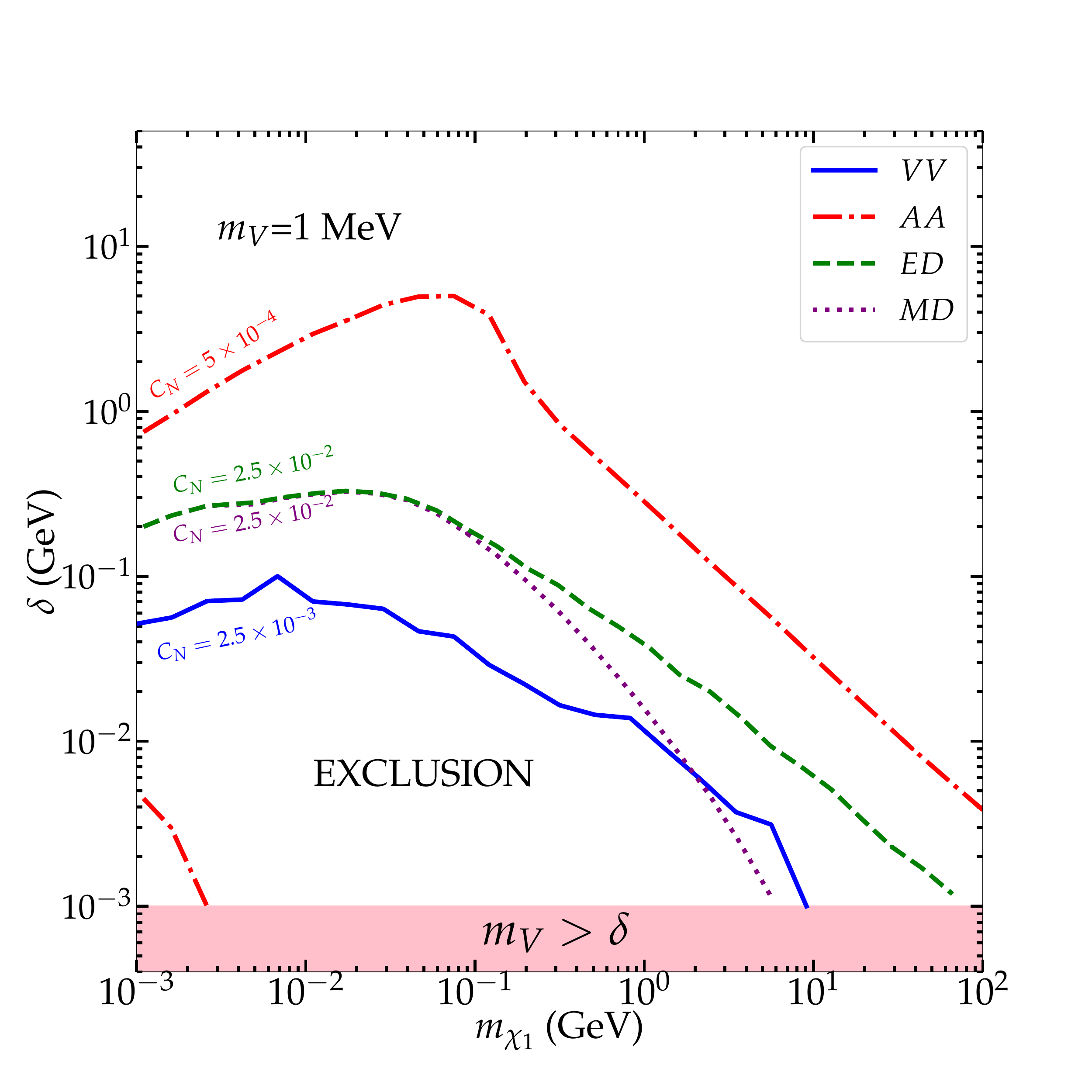}
\label{fig:delta_mx_c}
}
\subfloat[$m_{V}=10\mev$.]{
\includegraphics[width=0.49\textwidth]{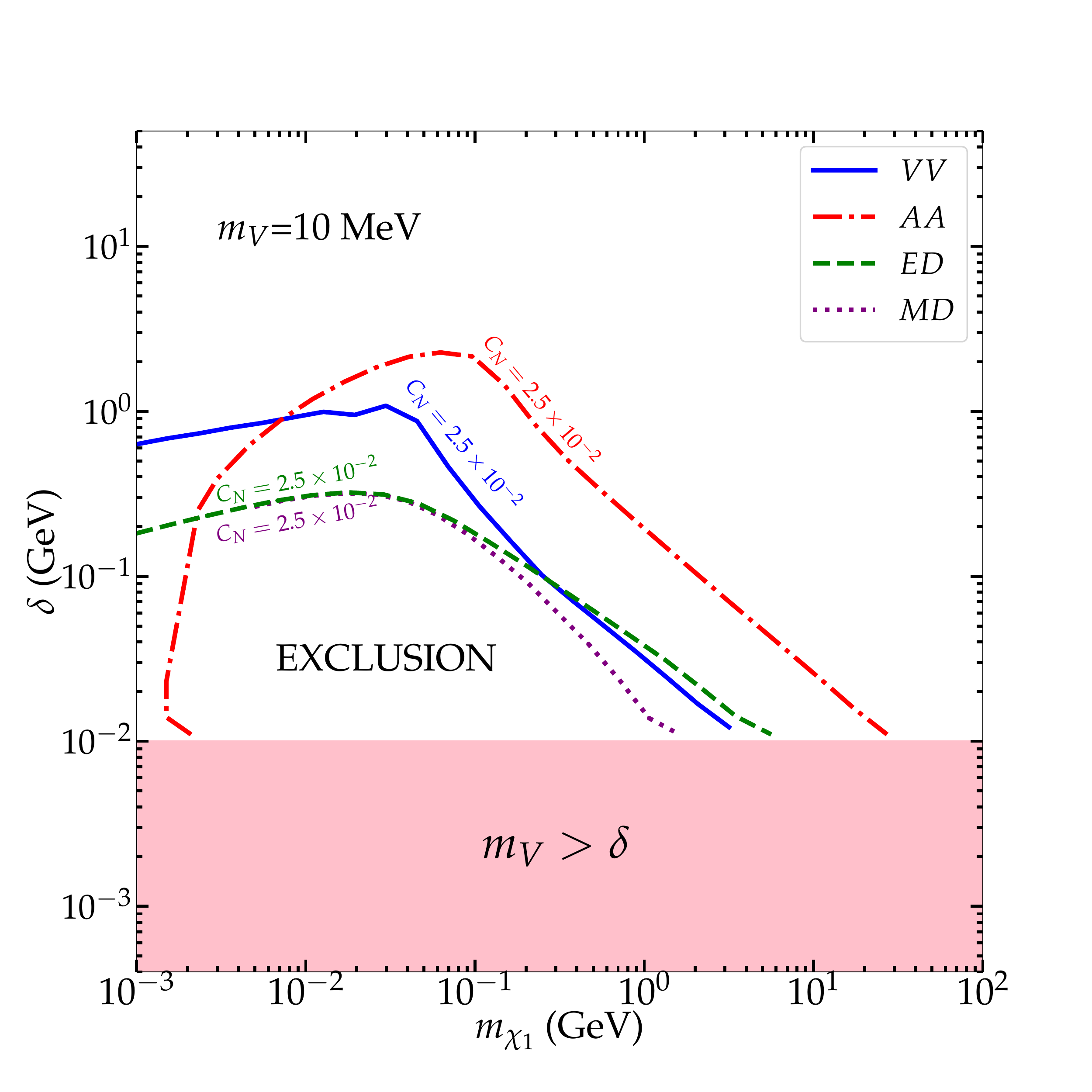}
\label{fig:delta_mx_d}
}
\caption{
PandaX-4T limits are projected onto the plane ($\delta$, $C_N$) [two upper panels] and
the plane ($m_{\chi_1}$, $\delta$) [two lower panels].
The line-of-sight halo integration $D_{\rm eff}$ is taken as 1~kpc.
The benchmark interactions are VV (blue solid lines), AA (red dash-dotted line), ED (green dashed line), and MD (purple dotted line).
In panel (a), ED and MD lines overlap.
We fixed $\Lambda_{E(M)}=100\mev$ for ED and MD.
}
\label{Fig:event_fixed_coupling}
\end{centering}
\end{figure*}

Results are shown in Fig.~\ref{Fig:event_fixed_coupling}.
Generally, there are four unknown variables that we are interested in:
$m_{\chi_1}$, $\delta$, $m_V$ and the coupling strength $C_N=C_N^{a/v}$.
Again, we take $D_{\rm eff}=1$~kpc, $C_\chi^v=C_\chi^a=1$ , and
$\Lambda_{E(M)}=100\mev$ for dipole-like interaction here.
As a reference, the $p \chi_1 \to p' \chi_2$ cross section with $C_N^{v/a} \approx 10^{-3}$ for VV and AA, respectively, and
$C^v_N \approx 10^{-2}$ for both MD and ED at a high $E_p$ are approximately equal to $\mathcal{O}(10^{-28})~\rm{cm}^2$.

In Fig.~\ref{fig:delta_CN_a} and ~\ref{fig:delta_CN_b},
by fixing $m_{\chi_1}$ and $m_{V}$, we obtain the limits of $C_N$ for each $\delta$
and the VV (blue solid line), AA (red dash-dotted line), ED (green dashed line), and
MD (purple dotted line) interactions.
An interesting feature is that the exclusion lines of $C_N$ remain relatively flat when
the mass splitting is less than a given value near $\delta\lesssim 10\mev$.
\textit{Thus, if $\delta$ is comparable to $m_V$, the total event rate is no longer sensitive to the changes of $\delta$. }
However, $\delta$ larger than $10\mev$ can weaken the limit.
As mentioned in the former sections, a larger $\delta$ can reduce the inelastic CRDM event rate.
Compared with the inelastic vDM scenario as the result from~\cite{PandaX-II:2017zex},
the inelastic CRDM helps us to probe a larger $\delta$ region.
When increasing $m_{\chi_1}$, the limits for the interactions with a $\gamma^5$ (AA and ED)
at the region $\delta<10^{-2}\gev$ are markedly improved,
but the limit for the MD interaction can be even weaker.
In the large $\delta$ region, PandaX-4T rapidly loses exclusion power if considering a large $m_{\chi_1}$.
Because the order of coupling strength can also represent the related size of the DM-proton cross section,
the PandaX-4T result projected on the AA interaction can give the most stringent limit on $C_N$
while the higher dimensional operators, especially MD, have weaker limits.

Fig.~\ref{fig:delta_mx_c} and~\ref{fig:delta_mx_d} show the exclusion limits from PandaX-4T projected to the ($m_{\chi_1}$, $\delta$) plane.
Due to the on-shell condition $\delta>m_V$, the pink region is not accessible.
Based on the information of Fig.~\ref{fig:delta_CN_a} and~\ref{fig:delta_CN_b},
we arbitrarily fix their couplings to optimize their detection as closer as
the PandaX-4T limit at $\delta\to 0$.
In Fig.~\ref{fig:delta_mx_c}, we find that only the exclusion region of the AA interaction is between two red dash-dotted lines,
while the exclusion regions are below the corresponding lines for other interactions.
These results occur because mass splitting (see Fig.~\ref{fig:Ep_Xsec_fer_cacv_a})
plays a role of enhancement in AA interaction.
The bottom left corner with $\delta< 5\times 10^{-2}\gev$ and $m_{\chi_1}<2\times 10^{-3}\gev$ becomes an allowed region again for AA interaction.
Thus, as long as we choose some stronger coupling,
this allowed region can sink to the inaccessible (pink) region.
In Fig.~\ref{fig:delta_mx_d}, we take all the coupling strengths $C_N=2.5\times 10^{-2}$.A turning point then appears in all the interactions, and
the limits behave differently at the smaller and larger $m_{\chi_1}$ regions.
We can understand these points and limits as follows.
The upper limits of $\delta$ at the large $m_{\chi_1}$ region decrease with respect to $m_{\chi_1}$, mainly because DM number density $\rho_0/m_{\chi_1}$ decreases.
The limits at the small $m_{\chi_1}$ region are similar to vDM limits, which always increase with $m_{\chi_1}$.
With heavy mediator masses $m_V=10\mev$ and $C_N=2.5\times 10^{-2}$, we find that PandaX-4T provides the most stringent limit for AA in the large $m_{\chi_1}$ region but for VV in the lower $m_{\chi_1}$ region.
With the help of the CRDM scenario, the PandaX-4T exclusion of $\delta$
can be extended to $0.1\gev$, even $1\gev$ in some cases.

\begin{figure*}[htbp]
\begin{centering}
\subfloat[$\delta=100\mev$]{
\includegraphics[width=0.49\textwidth]{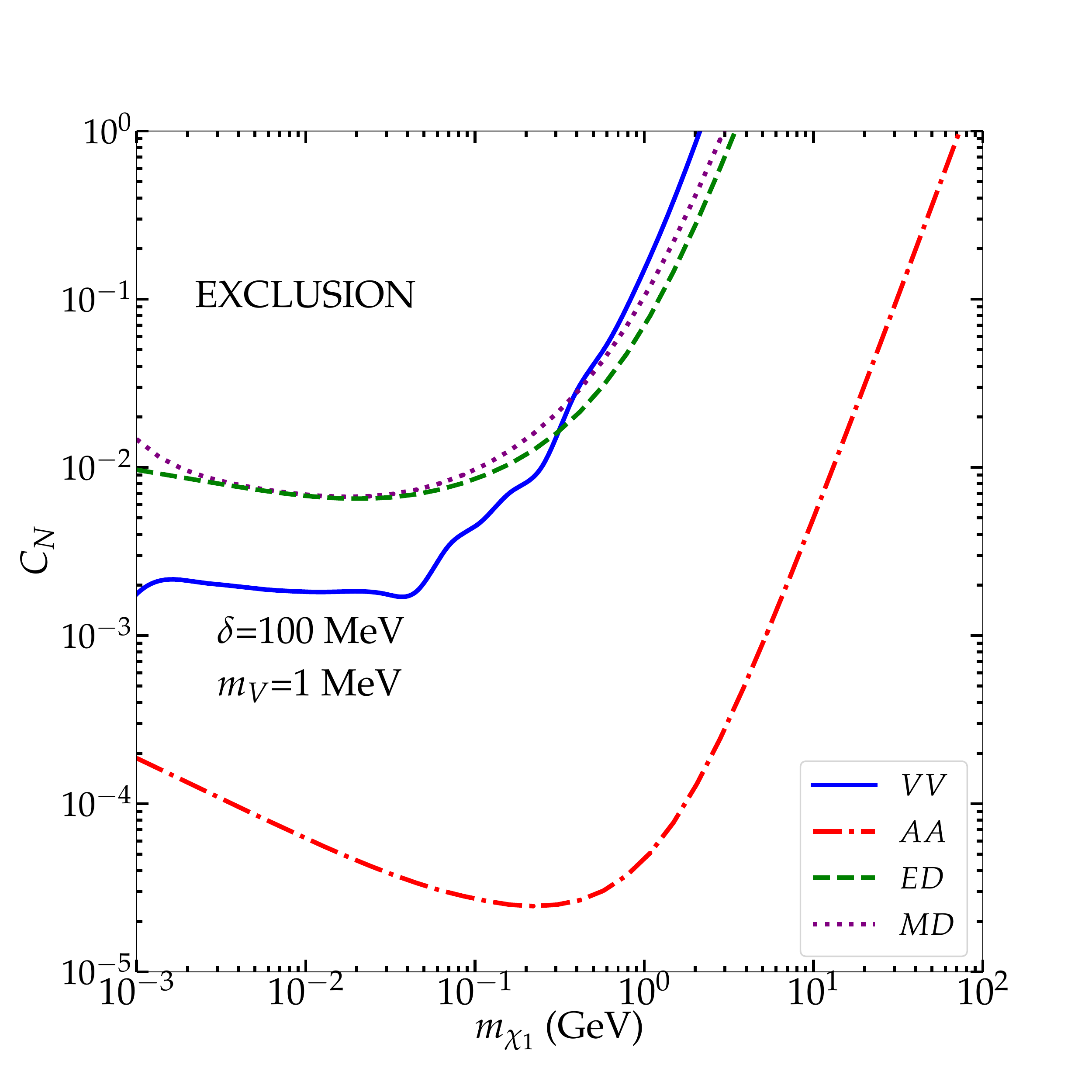}
\label{fig:mx_CN_a}
}
\subfloat[VV]{
\includegraphics[width=0.49\textwidth]{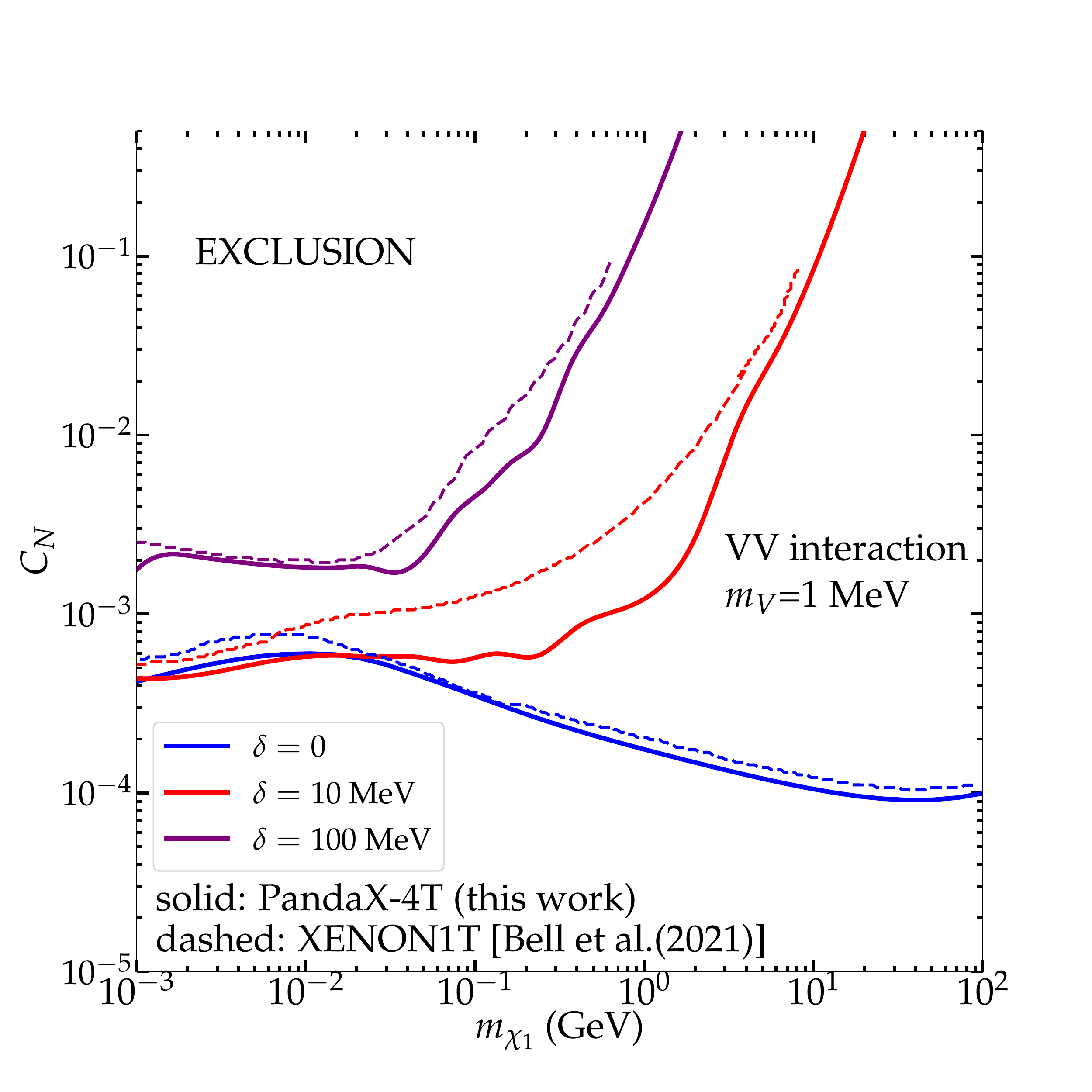}
\label{fig:mx_CN_b}
}
\caption{95\% upper limits projected in the ($m_\chi$, $C_N$) plane.
(a) the limits with fixed values $\delta=100\mev$ and $m_V=1\mev$ for
VV, AA, ED, and MD.
(b) Comparison between XENON1T and PandaX-4T based on the VV interaction.
The XENON1T limit from Ref.~\cite{Bell:2021xff} is used for comparison.
%In panel (a), we first compare our PandaX-4T constraints on VV interaction with XENON1T constraints from Ref.~\cite{Bell:2021xff}. Although we use different experimental data and techniques for the DM collision and decay, the results are on the same scale. In panel (b), The PandaX-4T constraints on different interaction types are projected onto the plane $(m_{\chi},C_N)$ with a fixed $\delta=100\mev$.
}
\label{Fig:event_fixed_delta}
\end{centering}
\end{figure*}

Finally, in Fig.~\ref{Fig:event_fixed_delta},
we present the $95\%$ upper limits projected in the ($m_\chi$, $C_N$) plane.
The mediator mass and astrophysical effective length are fixed to $m_V=1\mev$ and $D_{\rm eff}=1$~kpc, respectively.
In Fig.~\ref{fig:mx_CN_a}, the upper limits of PandaX-4T are based on the scenarios: VV (blue solid line),
AA (red dashed-dotted line), ED (green dashed line), and MD (purple dotted line).
With the fixed values $\delta=100\mev$,
the ordering of $95\%$ limits in four scenarios remain similar to those in Fig.~\ref{Fig:event_fixed_coupling}.
In Fig.~\ref{fig:mx_CN_b}, the previous XENON1T limits (dashed lines)
are taken from Ref.~\cite{Bell:2021xff}.
Because Ref.~\cite{Bell:2021xff} only considers the VV interaction,
we use the same parameter configurations as theirs for a comparison between
the elastic case (blue lines), $\delta=10\mev$ (red lines), and
$\delta=100\mev$ (purple lines).
Thus, the results of this study generally agree with the limits derived in Ref.~\cite{Bell:2021xff}.
We also notice that the CRDM fluxes are roughly $10$ orders of magnitudes lower than vDM, see 
Fig.~\ref{Fig:flux_fer1_a}. 
Because of the higher kinetic energy of CRDM,  
the PandaX-4T data can probe sub-GeV DM region in CRDM but not in vDM scheme. 
An example comparison is demonstrated in Fig.~\ref{fig:vdm_crdm} of the appendix~\ref{app:supp}.

\section{Conclusion and prospect}
\label{sec:conclusion}

In this study, we are motivated by the feature that DM relic density can be simply fulfilled at the co-annihilation region.
However, the DM located in this region escapes from the standard DM direct detection constraints.
Therefore, we investigated inelastic DM models
that contain a pair of almost mass-degenerated DM particles $\chi_1$ and $\chi_2$.
Because the standard DD method is not able to detect the vDM with extremely low momentum,
this kind of inelastic DM model can be hidden within the analysis unless the DM mass is heavy enough $m_{\chi_1}\sim\mathcal{O}(\tev)$.
By considering the relativistic CRDM events created by the collision between the nonrelativistic vDM
with the CR proton and helium, the lightest $\chi_1$ can be excited to $\chi_2$ and
successively decay back to $\chi_1$.
Energetic $\chi_1$ can be detected within the DM underground detector, such as PandaX-4T, in this study.
\textit{We have thus demonstrated that
mass splitting $\delta<\mathcal{O}(1\gev)$ can still be achieved with the DM mass range
considered in this study with the latest PandaX-4T data, even though we conservatively take the astrophysical parameter
$D_{\rm eff}=1$~kpc.}

Recently, a similar study~\cite{Bell:2021xff} considered the VV interaction and produced a rather flat cross section
with respect to the CR proton energy $E_p$. Beyond the scope of Ref.~\cite{Bell:2021xff}, we also studied several different interactions,
including both fermionic and scalar DM.
By studying the predicted CRDM spectra of the different interactions,
we found that the parameters ($m_{\chi_1}$, $\delta$, $m_V$, and $E_p$) play different roles in
the fermionic AA interactions of Eq.~\eqref{eq:f1}
compared with others, while the fermionic VV interaction of Eq.~\eqref{eq:f1} is almost identical
with the scalar interaction of Eq.~\eqref{eq:s1}.
Conversely, we also studied the dimension-suppressed dipole-like interactions of Eq.~\eqref{eq:f2} and Eq.~\eqref{eq:s2}.
Because $\chi_1$ can be relativistic before colliding with xenon in the detectors,
we used the velocity-dependent form factor based on the effective theory framework.
Therefore, our results for AA and dipole-like interactions are more reliable.

We then comment on possible constraints from other experiments that may be able to test the same parameter space.
We focused on the mass of mediator $m_V$ smaller than $\delta$ and $m_{\chi_1}$ in this study.
Then, we assumed that the mediator is leptophobic. With these two conditions, the cross sections of DM production via off-shell mediators in fixed targets, B-factories,
and LHC experiments are suppressed.
Therefore, if $C_N\lesssim 10^{-2}$, we can safely ignore the above constraints
in our scenarios\footnote{Note that this mass spectrum setting is different from
the usual ones with $m_V > 2m_{\chi_1}+\delta$ in Ref.~\cite{Izaguirre:2015zva,Berlin:2018jbm,Kang:2021oes}. }.
However, the light leptophobic mediator is confronted with the constraints of low-energy n-Pb scattering~\cite{Barbieri:1975xy,Leeb:1992qf}
and hadronic $\Upsilon (1S)$ decay~\cite{Aranda:1998fr}. 
In particular, the former provides the constraint $C_N\lesssim 5\times 10^{-3}$ for $m_V\lesssim 10\mev$~\cite{Tulin:2014tya}
but loses its exclusion power when $m_V$ increases.
We find that the upper limit of $C_N$ from the low energy n-Pb scattering can be complemented to the PandaX-4T limit derived in this study.
For a comparison, one can refer to Appendix~\ref{app:supp} for more details. 
More constraints and searches for light mediator $V$ can be found in Refs.~\cite{Tulin:2014tya,Chen:2019ivz,Suliga:2020lir}.

In the last paragraph, we would like to note several interesting extensions of this study,
even though they are beyond the scope of this study.
Interesting follow-up research should consider the Earth's attenuation effect of CRDM propagation.
In this study, we only focused on the exclusion limit, but the attenuation can be important
when the $\chi_1 p$ interaction cross section is large.
Compared with the scenario for elastic CRDM scattering with Earth atoms,
the inelastic scattering requires
two dark particle propagation equations.
In addition, the geometry of the propagation region might be complicated, and it can be challenging to solve
the propagation equations analytically.
Thus, we plan to return to this issue in the future with a numerical solution.
Another interesting follow-up study would describe the form factor more precisely.
In this study, we simply rescaled the form factor obtained by the assumption of elastic scattering.
In addition, only the leading order of the velocity contribution has been included.
Although much effort may be needed, this research would be useful, 
particularly if a more realistic form factor was developed for inelastic scattering.

%\newpage

\section*{Acknowledgments}
We would like to thank Jianglai Liu for providing the PandaX-4T efficiency table; Qian Yue and Jin Li for discussions about experimental issues; and Ran Ding, Gang Guo and Shao-Feng Ge for their valuable comments. This study was supported by the National Natural Science Foundation of China under Grant No. 11805012 and No. 12135004 and by the KIAS Individual Grant No. PG075302 at the Korea Institute for Advanced Study.

\appendix
\section{Kinematics of Two-body Inelastic collisions}
\label{app:kinematics}

As schematically shown in Fig.~\ref{Fig:scenario}, there are two scenarios of DM collisions:
(i) the high-energy protons of cosmic rays scattering with stationary DM $\chi$, namely, $p+\chi_1\rightarrow p'+\chi_2$,
where $\chi_2$ is the excited state of $\chi_1$ with a mass of $m_{\chi_2}=m_{\chi_1}+\delta$;
(ii) the accelerated DM colliding with the stationary nucleus $N$ in the detector $\chi_1+N\rightarrow \chi_2+N' $.

Considering the general two-body relativistic collision, $p_1 + p_2 \rightarrow p_3 + p_4$,
their masses are $m_{i}$ with $i=1,2,3,4$.
We can write down the 4-momentum of each particle of the process in lab frame $\Sigma^L$:
\begin{eqnarray}
&& p_1=(E_1,{\bf p}_1),
\nonumber\\
&& p_2=(E_2=m_2, {\bf p}_2\simeq 0),
\nonumber\\
&& p_3=(E_3,{\bf p}_3),
\nonumber\\
&& p_4=(E_4,{\bf p}_4).
\label{eq:lab_frame}
\end{eqnarray}
In the center-of-mass (CM) frame $\Sigma^*$, all physical quantities are marked with $*$:
\begin{eqnarray}
&& p^*_1=(E^*_1,{\bf p}^*_1),
\nonumber\\
&& p^*_2=(E^*_2,{\bf p}^*_2=-{\bf p}^*_1),
\nonumber\\
&& p^*_3=(E^*_3,{\bf p}^*_3),
\nonumber\\
&& p^*_4=(E^*_4,{\bf p}^*_4=-{\bf p}^*_3),
\label{eq:CM_frame}
\end{eqnarray}
The total 4-momentum $P$ of two systems are simply:
 \begin{eqnarray}
P(\Sigma^*)=(M,0),~~{\rm and}~~P(\Sigma^L)= (E_1+m_2,{\bf p}_1).
\end{eqnarray}
Also:
\begin{eqnarray}
|{\bf p}^*_3|&=&\frac{1}{2} \sqrt{\frac{\left(m_3^2-m_4^2\right)^2}{M^2}+M^2-2
   \left(m_3^2+m_4^2\right)}, ~~{\rm and} \nonumber\\
E^*_4 &=& \frac{M^2-m_3^2+m_4^2}{2 M}.
\label{eq:p3_E4}
\end{eqnarray}
Boosting from the CM frame to the lab frame, one will obtain the magnitude of velocity $\beta=\left|{\bf p}_1\right|/\left(E_1+m_{2}\right)$ and
Lorentz factor $\gamma=(E_1+m_2)/M$. Therefore,
the invariant mass $M$ will be related to $E_1$ via:
\begin{eqnarray}
M=(E_1+m_2) \sqrt{1-\beta^2}=
\sqrt{m_{2}^2+m_1^2+2m_{2}E_1}.
\label{eq:invM}
\end{eqnarray}

\subsection{Accelerating process $p+\chi_1\rightarrow p'+\chi_2$}
\label{app:px1px2}
In the process of $p(p_1)+\chi_1(p_2)\rightarrow p'(p_3)+\chi_2(p_4)$, where the four-momentum label has been given in parentheses, one has $p_2=(E_2=m_{\chi_1}, {\bf p}_2\simeq 0)$, and $\chi_2$ is the accelerated DM after collision, with energy $E_4$.
We can insert $E_1=E_p$, $m_1=m_3=m_p$, $m_2=m_{\chi_1}$, and $m_4=m_{\chi_1}+\delta$ into Eq.~\eqref{eq:p3_E4} and obtain:
\begin{eqnarray}
E_{4}^*=E_{\chi_2}^*=&&\frac{m_{\chi_1}(m_{\chi_1}+E_p)+\delta(\delta/2+m_{\chi_1})}{M}.
\label{eq:Estar4}
\end{eqnarray}
If $\delta=0$, we can safely return to the case of elastic collision.
Also, the condition $E_{\chi_2}^*>(m_{\chi_1}+\delta)$ is required
for $\chi_1$ being excited to $\chi_2$. This results in a universal lower limit for $E_p$, which is $m_p + \delta + \frac{\delta(\delta+2 m_p)}{2m_{\chi_1}}$.

To determine the minimum value of $E_{p}$ in Eq.~\eqref{eq:x1flx},
it would be useful to introduce the scattering angle $\theta^*$ where
the component of ${\bf p}_3^*$ along the direction of ${\bf p}_1^*$
is $|{\bf p}^*_3|\cos\theta^*$.
Therefore, we substitute $E_4$ by $T_{\chi_2}+m_{\chi_1}+\delta$ to obtain:
\begin{eqnarray}
T_{\chi_2}=&&\frac{E_{\chi_2}^*(m_{\chi_1}+E_p)-{|\bf p}^*_{\chi_2}|\sqrt{E_p^2-m_p^2}\cos\theta^*}{\sqrt{(m_{\chi_1}+m_p)^2+2m_{\chi_1}(E_p-m_p)}}
-(m_{\chi_1}+\delta).
\label{eq:Tchi2}
\end{eqnarray}
where {$|{\bf p}^*_{\chi_2}| = \sqrt{(E^*_{\chi_2})^2-(m_{\chi_1}+\delta)^2}$.}

\begin{figure}[htbp]
 \centering
    \includegraphics[width=0.5\textwidth]{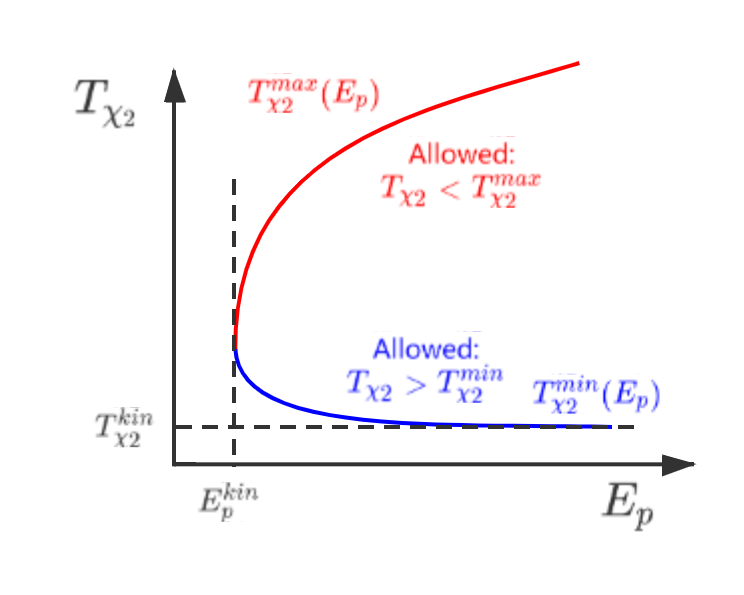}
 \caption{Kinetic relation between incoming $E_p$ and outgoing $T_{\chi_2}$. From Eq.~\eqref{eq:Tchi2}, the red curve represents $T_{\chi_2}^{max}$ with $\theta^*=\pi$, and the blue curve represents $T_{\chi_2}^{min}$ with $\theta^*=0$. The kinetic allowed region is between the red and blue curves. The two curves can be uniformly described by Eq.~\eqref{eq:Epmin}. Two limits of the curve correspond to two kinematic constraints for inelastic collision: $T_{\chi_2}^{kin} = \frac{\delta^2}{2m_{\chi_1}}$ and $E^{kin}_p = m_p + \delta + \frac{\delta(\delta+2 m_p)}{2m_{\chi_1}}$.}
    \label{Fig:Epmin}
\end{figure}

We would like to obtain the range of $E_p$ from Eq.~\eqref{eq:Tchi2}. Because $T_{\chi_2}(E_p,\cos{\theta^*})$ reaches maximum and minimum values when $\theta^*=\pi$ and $\theta^*=0$, respectively, we may define $T_{\chi_2}^{max}(E_p)=T_{\chi_2}(E_P,\theta^*=\pi)$ and $T_{\chi_2}^{min}(E_p)=T_{\chi_2}(E_P,\theta^*=0)$.
For a fixed $E_p$, the allowed region of $T_{\chi_2}$ should be between $T_{\chi_2}^{min}$ and $T_{\chi_2}^{max}$, illustrated in Fig.~\ref{Fig:Epmin}.

Similar to the elastic case, the constraint of $T_{\chi_2}$ can be expressed by $E_p^{min}(T_{\chi_2})$:

\begin{eqnarray}
E_p>E^{\rm min}_p=\frac{T_{\chi_2}+\delta}{2}
+
\sqrt{
\frac{
T_{\chi_2}(m_{\chi_1}+\frac{T_{\chi_2}}{2}+\delta)(2m_p^2+m_{\chi_1}T_{\chi_2}-\frac{\delta^2}{2})}
{2m_{\chi_1}T_{\chi_2}-\delta^2}
}.
\label{eq:Epmin}
\end{eqnarray}
where $2m_{\chi_1}T_{\chi_2}>\delta^2$ should be satisfied. The minimum value of $E^{min}_p$ is $m_p + \delta + \frac{\delta(\delta+2 m_p)}{2m_{\chi_1}}$. The condition $E_p>E_p^{min}(T_{\chi_2})$ is equivalent to $T_{\chi_2}^{min}(E_p)<T_{\chi_2}<T_{\chi_2}^{max}(E_p)$.

\subsection{Process in the detector $\chi_1+N\to \chi_2+N'$}

For the process $\chi_1(p_1)+N(p_2)\rightarrow \chi_2(p_3)+N'(p_4)$ with $p_2=(E_2=m_N, {\bf p}_2\simeq 0)$,
we can swap the index that the incoming particle $1$ is inelastic DM,
but $E_4$ is the energy of the heavy nucleus after the collision.
By taking the masses $m_2=m_4=m_N$, $m_1=m_{\chi_1}$, and $m_3=m_{\chi_1}+\delta$,
we can rewrite Eq.~\eqref{eq:p3_E4} as:
\begin{eqnarray}
E^*_4=E^*_{N'}=\frac{m_N(m_N+E_{\chi})-\delta(\delta/2+m_{\chi_1})}{M}.
\label{eq:Estar4_det}
\end{eqnarray}

We can boost Eq.~\eqref{eq:Estar4_det} from the CM frame to the Lab frame to obtain
the recoil energy $Q=E_4-m_N$ of nucleus.
Finally, the minimal kinetic energy required to obtain a specific recoil energy $Q$ is:
\begin{eqnarray}
T^{min}_{\chi}=&&
\frac{Q}{2}-m_{\chi_1}+\frac{\delta(m_{\chi_1}+\delta/2)}{2m_N}\nonumber \\
&&+\frac{\sqrt{Q(2m_N+Q)(m_N Q+\delta^2/2)(m_N Q+(2m_{\chi_1}+\delta)^2/2)}}
{2m_N Q}.
\label{eq:Txmin}
\end{eqnarray}

\section{Scattering Cross Sections}
\label{app:Xsec}

\subsection{$2\to 3$ cross sections}
The differential cross section for the inelastic scattering $p(p_1) \chi_1(p_2) \rightarrow p'(k_1) \chi_1(k_2) V(k_3)$ can be represented as:
\begin{eqnarray}
d\sigma_{p\chi_1\rightarrow p'\chi_1 V}=\frac{(2\pi)^4 |\mathcal{M}_{2\rightarrow3}|^2}{4\sqrt{(p_1\cdot p_2)^2-m_p^2m_{\chi_1}^2}}d\phi_3(p_1+p_2;k_1,k_2,k_3),
\label{eqn:master}
\end{eqnarray}
where $\mathcal{M}_{2\rightarrow3}$ is the total scattering amplitude, and $\phi_3(p_1+p_2;k_1,k_2,k_3)$ is the three-body phase space. Using the recursive relation, we have:
\begin{eqnarray}
d\phi_3(p_1+p_2;k_1,k_2,k_3)= d\phi_2(p_1+p_2;k_1,q)\times d\phi_2(q;k_2,k_3)(2\pi)^3dq^2
\label{eqn:phase_decom}
\end{eqnarray}
where $q$ is the four-momentum of $\chi_2$. This kind of technique is extensively used in Ref.~\cite{Kang:2016jxw}
With the narrow width approximation, which corresponds to the case of on-shell $\chi_2$, we can separate the scattering process into collision and decay parts, namely, $p \chi_1 \rightarrow p' \chi_2$ and $\chi_2 \rightarrow \chi_1 V$.
Then, $|\mathcal{M}_{2\rightarrow3}|^2$ can be written as:
\begin{eqnarray}
|\mathcal{M}_{2\rightarrow3}|^2=
|\mathcal{M}_{2\rightarrow2}|^2 \frac{\pi\delta(q^2-m_{\chi_2}^2)}{m_{\chi_2}\Gamma_{\chi_2}}
|\mathcal{M}_{1\rightarrow 2}|^2,
\label{eqn:amp_decom}
\end{eqnarray}
where $\mathcal{M}_{2\rightarrow2}$ is the scattering amplitude of the collision part, and $\mathcal{M}_{1\rightarrow 2}$ is that of the decay part.

We can decompose the total differential cross in the following form:
\begin{eqnarray}
d\sigma_{p\chi_1\rightarrow p'\chi_1 V}=d\sigma_{p\chi_1 \rightarrow p' \chi_2}dB_{\chi_2 \rightarrow \chi_1 V},
\label{eqn:cro_decom}
\end{eqnarray}
where we have:
\begin{eqnarray}
d\sigma_{p\chi_1 \rightarrow p' \chi_2}=
\frac{(2\pi)^4 |\mathcal{M}_{2\rightarrow2}|^2}{4\sqrt{(p_1\cdot p_2)^2-m_p^2m_{\chi_1}^2}}
d\phi_2(p_1+p_2;k_1,q),
\label{eqn:coll}
\end{eqnarray}
And:
\begin{eqnarray}
dB_{\chi_2 \rightarrow \chi_1 X}&&=\frac{2\pi}{\Gamma_{\chi_2}}\delta(q^2-m_{\chi_2}^2)
\frac{|\mathcal{M}_{1\rightarrow 2}|^2}{2m_{\chi_2}}d\phi_2(q;k_2,k_3)(2\pi)^3 dq^2
\nonumber\\
&&=\frac{d\Gamma_{\chi_2\rightarrow \chi_1 V}}{\Gamma_{\chi_2}}\delta(q^2-m_{\chi_2}^2)dq^2.
\label{eqn:decay}
\end{eqnarray}
The total width of $\chi_2$ is defined as $\Gamma_{\chi_2}$.

Finally, we successfully obtain the total differential cross section as Eq.~\eqref{eqn:dcro_total0}:
\begin{eqnarray}
\frac{d\sigma_{p\chi_1\rightarrow p'\chi_1 V}}{dT_{\chi_1}}=\int \frac{d\sigma_{p\chi_1\rightarrow p'\chi_2}}{dT_{\chi_2}}
\frac{dT_{\chi_2}}{dT_{\chi_1}}
\frac{dB_{\chi_2 \rightarrow \chi_1 V}}{d\cos{\theta'}} d\cos{\theta'},
\label{eqn:dcro_total}
\end{eqnarray}
where $\theta'$ is the angle of $\chi_1$ in the $\chi_2$ rest frame. In this frame:
\begin{eqnarray}
\frac{d\Gamma_{\chi_2 \rightarrow \chi_1 V}}{d\Omega'}=
\frac{|p^*_{\chi_1}|}{32\pi^2 q^2} |\mathcal{M}_{1 \rightarrow 2}|^2
\end{eqnarray}
where $\Omega'$ is the solid angle. In the presence of vector and axial vector interactions, we can prove that $|\mathcal{M}_{1 \rightarrow 2}|^2$ is independent of the angle. Then, the aforementioned $\theta'$ can be chosen as the angle formed by $\chi_2$ momentum in the lab frame of $p\chi_1$ and $\chi_1$ momentum for $\chi_2\to\chi_1 V$ decay in the $\chi_2$ rest frame. Therefore:
\begin{eqnarray}
dB_{\chi_2 \rightarrow \chi_1 V}=\frac{d\Gamma_{\chi_2\rightarrow \chi_1 V}}{\int d\Gamma_{\chi_2\rightarrow \chi_1 V} }\delta(q^2-m_{\chi_2}^2)dq^2\longrightarrow\frac{d\Omega'}{4\pi}
\end{eqnarray}
Ultimately, the factor of $\delta(q^2-m_{\chi_2}^2) dq^2$ will drop out in the process of integration over the three-body phase space \cite{Kang:2013jaa,Zhang:2020dla}. A trivial integration over $d\phi'$ gives the $dB_{\chi_2 \rightarrow \chi_1 V}/d\cos{\theta'} = 1/2$.

\subsection{ $2\to 2$ cross sections}
\label{app:Xsec2to2}
\subsubsection{The accelerating process $p+\chi_1\rightarrow p'+\chi_2$}
To calculate the total differential cross section between cosmic ray protons and inelastic dark matter in Eq.~\eqref{eqn:dcro_total}, we must derive $d\sigma_{p\chi_1\rightarrow p'\chi_2} / dT_{\chi_2} $ first:
\begin{eqnarray}
\frac{d\sigma_{p\chi_1\rightarrow p'\chi_2}}{dT_{\chi_2}}=\frac{d\sigma_{p\chi_1\rightarrow p'\chi_2}}{dt}\left|\frac{dt}{dT_{\chi_2}}\right|
=\frac{\overline{|\mathcal{M}|^2}}{16\pi\lambda(s,m_p^2,m_{\chi_1}^2)}\left|\frac{dt}{dT_{\chi_2}}\right|
\label{dsigma_dtx2}
\end{eqnarray}
where $\overline{|\mathcal{M}|^2}=\sum_{spins} |\mathcal{M}|^2/4$ is the proton-DM scattering matrix element squared, averaged over initial spins and summed over final spins. We define the Kallen function as $\lambda(x,y,z)=(x-y-z)^2-4yz$, and the Mandelstam variables are:
\begin{eqnarray}
&&s=m_{\chi_1}^2+m_p^2+2m_{\chi_1}E_p,
\nonumber\\
&&t=-2m_{\chi_1}T_{\chi_2}+\delta^2,
\nonumber\\
&&u=m_p^2+m_{\chi_1}^2-2m_{\chi_1}(E_p-T_{\chi_2}-\delta).
\label{mandel_CRDM}
\end{eqnarray}

%Using above variables, one can write down the %$d\sigma_{p\chi_1\to p'\chi_2} / dT_{\chi_2} $ of %scattering between CR proton and inelastic DM.
The amplitude squared $|\mathcal{M}_{ij}|^2$ with $i$ for the $VNN$ interaction type and
$j$ for the $V\chi\chi$ interaction type are given below.

For fermionic DM:
\begin{itemize}
\item Vector-Vector interaction:
\begin{eqnarray}
\overline{|\mathcal{M}_{\text{VV}}|^2}&=&
\left[ \frac{4 (C_N^v)^2 (C_{\chi}^v)^2 m_{\chi_1 }
}
{\left(2 m_{\chi_1 } T_{\chi_2}+m_{V }^2 -\delta ^2\right)^2}
\right]\times
\Big[-4 m_{\chi _1} E_p \left(\delta +T_{\chi _2}\right)+4 m_{\chi _1} E_p{}^2
\nonumber\\
&&-2 T_{\chi _2} \left(m_p^2+m_{\chi _1} \left(m_{\chi _1}-T_{\chi _2}\right)\right)+\delta^2 \left(m_{\chi _1}-T_{\chi _2}\right)\Big].
\label{vv_int_CRDM}
\end{eqnarray}

\item Axial vector-Axial vector interaction:
\begin{eqnarray}
\overline{|\mathcal{M}_{\text{AA}}|^2}&=&\left[ \frac{4 (C_N^a)^2 (C_{\chi}^a)^2 m_{\chi_1 } }
{\left(2 m_{\chi_1 } T_{\chi_2}+m_{V }^2 -\delta ^2\right)^2}
\right]\times\Big[-4 m_{\chi _1} E_p \left(\delta +T_{\chi _2}\right)+4 m_{\chi_1} E_p^2 \nonumber\\
&& +2 m_p^2 \left(4 \delta
   +T_{\chi _2}\right)+8 m_p^2 m_{\chi _1}
   +m_{\chi _1} \left(-\delta ^2+4 \delta  T_{\chi _2}+2 T_{\chi_2}^2\right) \nonumber\\
&& +2 m_{\chi _1}^2 T_{\chi _2}+\delta ^2
  \left(-\left(2 \delta +T_{\chi_2}\right)\right)\nonumber\\
&&+(2 m_p^2 T_{\chi _2} \left(\delta +2 m_{\chi _1}\right){}^2 \left(-\delta ^2+2 m_{\chi _1} T_{\chi_2}+2 m_V^2\right))/m_V^4\Big].
\label{aa_int_CRDM}
\end{eqnarray}

\item Vector-Magnetic Dipole interaction:
\begin{eqnarray}
\overline{|\mathcal{M_{\text{MD}}}|^2}&=&\left[ \frac{4 (C_N^v)^2m_{\chi_1}/(\Lambda_M)^2}
{\left(2 m_{\chi_1 } T_{\chi_2}+m_{V }^2 -\delta ^2\right)^2}
\right]\times4\Big[
4 m_{\chi _1} E_p^2 \left(2 m_{\chi _1} T_{\chi _2}-\delta ^2\right)\nonumber\\
&&-4 m_{\chi _1} E_p \left(\delta +T_{\chi _2}\right) \left(2 m_{\chi _1} T_{\chi _2}-\delta ^2\right) \nonumber\\&&-2 m_p^2 T_{\chi _2} \left(\delta +2 m_{\chi _1}\right){}^2+\left(2 m_{\chi _1} T_{\chi _2}-\delta ^2\right)\nonumber\\&& \times\left(\delta^2 \left(m_{\chi _1}+T_{\chi _2}\right)+4 \delta  m_{\chi _1} T_{\chi _2}+2 m_{\chi _1}^2 T_{\chi_2}\right)\Big].
\label{md_int_CRDM}
\end{eqnarray}

\item Vector-Electric Dipole interaction:
\begin{eqnarray}
\overline{|\mathcal{M}_{\text{ED}}|^2}&=&\left[ \frac{4 (C_N^v)^2m_{\chi_1}/(\Lambda_E)^2}
{\left(2 m_{\chi_1 } T_{\chi_2}+m_{V }^2 -\delta ^2\right)^2}
\right]\times4[4 m_{\chi _1} E_p{}^2 \left(2 m_{\chi _1} T_{\chi _2}-\delta ^2\right)\nonumber\\
&&-4 m_{\chi _1} E_p \left(\delta+T_{\chi _2}\right) \left(2 m_{\chi _1} T_{\chi_2}-\delta^2\right)\nonumber\\
&&-\delta ^2 m_{\chi _1} \left(3\delta ^2+4 m_p^2-4 \delta  T_{\chi _2}-2 T_{\chi _2}^2\right)\nonumber\\
&&-\delta ^2 \left(\delta ^2+2
m_p^2\right) \left(2 \delta +T_{\chi_2}\right)+8\delta^2 m_{\chi _1}^2 T_{\chi _2}-4 m_{\chi _1}^3T_{\chi _2}^2\Big].
\label{ed_int_CRDM}
\end{eqnarray}

For scalar DM:
\item Combining the vector interaction for SM and $\mathcal{L}^s_1$ in Eq.~\eqref{eq:s1} for scalar DM yields:
\begin{eqnarray}
\overline{|\mathcal{M}|^2}&=&\left[\frac{4 (C_N^v)^2 (g_\chi)^2 m_{\chi_1 } }
{\left(2 m_{\chi_1 } T_{\chi_2}+m_{V }^2 -\delta ^2\right)^2}
\right]\nonumber\\
&&\times\Big[m_{\chi _1} \left(4 E_p{}^2+\delta ^2- 4 E_p \left(\delta +T_{\chi _2}\right)-2 m_{\chi _1} T_{\chi_2}\right)\Big].
\label{sc1_int_CRDM}
\end{eqnarray}

\item Combining the vector interaction for SM and $\mathcal{L}^s_2$ in Eq.~\eqref{eq:s2} for scalar DM yields:
\begin{eqnarray}
\overline{|\mathcal{M}|^2}&=&\left[\frac{4 (C_N^v)^2 m_{\chi_1}/(\Lambda_s)^4}
{\left(2 m_{\chi_1 } T_{\chi_2}+m_{V }^2 -\delta^2\right)^2}\right]\nonumber\\
&&\times\frac{1}{4}\Big[m_{\chi _1} \left(4 E_p{}^2+\delta^2
-4 E_p \left(\delta +T_{\chi _2}\right)-2 m_{\chi _1} T_{\chi_2}\right)
\left(2 m_{\chi _1} T_{\chi _2}-\delta ^2\right)^2\Big].
\label{sc2_int_CRDM}
\end{eqnarray}

\end{itemize}

The corresponding differential cross section can be written as:
\begin{eqnarray}
\frac{d\sigma_{p\chi_1\rightarrow p'\chi_2}}{dT_{\chi_2}}=
\frac{m_{\chi_1}}{8\pi\lambda(s,m_p^2,m_{\chi_1}^2)}\overline{|\mathcal{M}|^2}
\label{eqn:dcroCRDM_22}
\end{eqnarray}

\subsubsection{Process in the detector $\chi_1+N\to \chi_2+N'$}
Similarly, we can derive $d\sigma_{\chi N}/d Q$ of Eq.~\eqref{eq:R_rel}.
To calculate $\overline{|\mathcal{M}|^2}$, we must only change the Mandelstam variables as:
\begin{eqnarray}
&&s=m_{\chi_1}^2+m_N^2+2m_N(T_{\chi_1}+m_{\chi_1}),
\nonumber\\
&&t=-2 m_N Q,
\nonumber\\
&&u=(m_N-m_{\chi_1})^2+2m_N(Q-T_{\chi_1})+\delta(2m_{\chi_1}+\delta).
\label{mandel_detect}
\end{eqnarray}
The amplitude squared $|\mathcal{M}_{ij}|^2$ with $i$ for the $Vpp$ interaction type and
$j$ for the $V\chi\chi$ interaction type are given below.

For fermionic DM:
\begin{itemize}
\item Vector-Vector interaction:
\begin{eqnarray}
\overline{|\mathcal{M}_{\text{VV}}|^2}&=&
\left[\frac{4 (C_N^v)^2 (C_\chi^v)^2 m_N}{\left(2 Q m_N+m_V^2\right)^2}\right]\times
\Big[m_N \left(-\delta ^2-4 \left(m_{\chi _1}+T_{\chi _1}\right) \left(-m_{\chi _1}+Q-T_{\chi _1}\right)+2Q^2\right)\nonumber\\
 &&-2 Q m_N^2-2 m_{\chi _1} \left(\delta ^2+m_{\chi _1} (2 \delta +Q)\right)-2 \delta  T_{\chi
   _1} \left(\delta +2 m_{\chi _1}\right)+\delta ^2 Q\Big]
\label{vv_int_det}
\end{eqnarray}

\item Axial vector-Axial vector interaction:
\begin{eqnarray}
\overline{|\mathcal{M}_{\text{AA}}|^2}&=&
\left[\frac{4 (C_N^a)^2 (C_\chi^a)^2 m_N}{\left(2 Q m_N+m_V^2\right)^2}\right]\nonumber\\
&&\times\Big[m_N \left(\delta ^2+m_{\chi _1} \left(8 \delta -4 Q+8 T_{\chi _1}\right)+12 m_{\chi _1}^2+2 Q^2+4 T_{\chi1} \left(T_{\chi_1}-Q\right)\right)\nonumber\\
&&+2 Q m_N^2+2 m_{\chi _1}^2 (Q-2 \delta )-2 \delta  m_{\chi_1}\left(\delta -2 Q+2 T_{\chi _1}\right)+\delta^2 \left(Q-2 T_{\chi _1}\right)\nonumber\\
&& +\left(2 m_N \left(\delta +2 m_{\chi _1}\right){}^2 \left(2 Q^2 m_N^2+Q m_N \left(\delta ^2+2 m_V^2\right)+\delta ^2 m_V^2\right)\right)/m_V^4\Big].
\label{aa_int_det}
\end{eqnarray}

\item Vector-Magnetic Dipole interaction:
\begin{eqnarray}
\overline{|\mathcal{M}_{\text{MD}}|^2}&=&\left[\frac{4 (C_N^v)^2 m_N/(\Lambda_M)^2}{\left(2 Q m_N+m_V^2\right)^2}\right]\times4\Big[4 m_N m_{\chi _1}^2 \left(-\delta ^2+Q^2-2 \delta  Q\right)\nonumber\\
&&-\delta ^2 m_N \left(\delta ^2-2 Q^2+4 Q T_{\chi_1}\right)-4 \delta  m_N m_{\chi _1} \left(\delta ^2-2 Q^2+\delta Q+2 Q T_{\chi_1}\right)\nonumber\\
&&+\delta ^2 Q \left(\delta +2 m_{\chi_1}\right)^2-2 Q m_N^2 \left(\delta ^2+4 m_{\chi _1} \left(\delta +Q-2 T_{\chi _1}\right)+4 Q T_{\chi _1}-4 T_{\chi_1}^2\right)\Big].\nonumber\\
\label{md_int_det}
\end{eqnarray}

\item Vector-Electric Dipole interaction:
\begin{eqnarray}
\overline{|\mathcal{M}_{\text{ED}}|^2}&=&\left[\frac{4 (C_N^v)^2 m_N/(\Lambda_E)^2}{\left(2 Q m_N+m_V^2\right)^2}\right]\times4\Big[-\delta ^2 m_N \left(\delta ^2-2 Q^2+4 Q T_{\chi _1}\right)-4 m_N m_{\chi_1}^2 (\delta +Q)^2\nonumber\\
&&-4 \delta  m_N m_{\chi _1} \left(\delta ^2+\delta  Q+2 Q T_{\chi _1}\right)+\delta ^2 Q \left(\delta +2 m_{\chi _1}\right)^2\nonumber\\
&&-2 Q m_N^2 \left(\delta ^2+4 m_{\chi
   _1} \left(Q-2 T_{\chi _1}\right)-4 m_{\chi _1}^2+4 Q T_{\chi _1}-4 T_{\chi _1}^2\right)\Big].
\label{ed_int_det}
\end{eqnarray}

For scalar DM:

\item Combining the vector interaction for SM and $\mathcal{L}^s_1$ in Eq.~\eqref{eq:s1} for scalar DM yields:
\begin{eqnarray}
\overline{|\mathcal{M}|^2}&=&\left[\frac{4 (C_N^v)^2 (g_\chi)^2m_N}{\left(2 Q m_N+m_V^2\right)^2}\right]
\times2\Big[m_{\chi_1}^2 \left(-2 \delta +2 m_N-Q\right) \nonumber\\
&&+m_{\chi_1} \left(-\delta ^2-2 Q m_N+2 T_{\chi _1} \left(2   m_N-\delta \right)\right)+T_{\chi _1} \left(2 m_N \left(T_{\chi _1}-Q\right)-\delta ^2\right)\Big].\nonumber\\
\label{sc1_int_det}
\end{eqnarray}

\item Combining the vector interaction for SM and $\mathcal{L}^s_2$ in Eq.~\eqref{eq:s2} for scalar DM yields:
\begin{eqnarray}
\overline{|\mathcal{M}|^2}&=&\left[\frac{4 (C_N^v)^2m_N/(\Lambda_s)^4}{\left(2 Q m_N+m_V^2\right)^2}\right]
\times2 Q^2 m_N^2\Big[2 m_N\left(m_{\chi_1}+T_{\chi_1}\right) \left(m_{\chi_1}-Q+T_{\chi_1}\right)\nonumber\\
&&-m_{\chi _1}^2 (2 \delta +Q)-\delta  m_{\chi_1} \left(\delta +2 T_{\chi _1}\right)-\delta^2 T_{\chi _1}\Big].
\label{sc2_int_det}
\end{eqnarray}

\end{itemize}
Again, the corresponding differential cross section is
\begin{eqnarray}
\frac{d\sigma_{\chi N}}{dQ}=
\frac{m_{N}}{8\pi\lambda(s,m_{\chi_1}^2,m_N^2)}\overline{|\mathcal{M}|^2}.
\label{dcrosec_detect}
\end{eqnarray}

\section{Supplemental figures}
\label{app:supp}

In this section, we compare the new limits derived in this work with previous constraints, 
to demonstrate the advantage of CRDM for inelastic DM study.

\begin{figure*}[htbp]
\begin{centering}
\subfloat[]{
\includegraphics[width=0.49\textwidth]{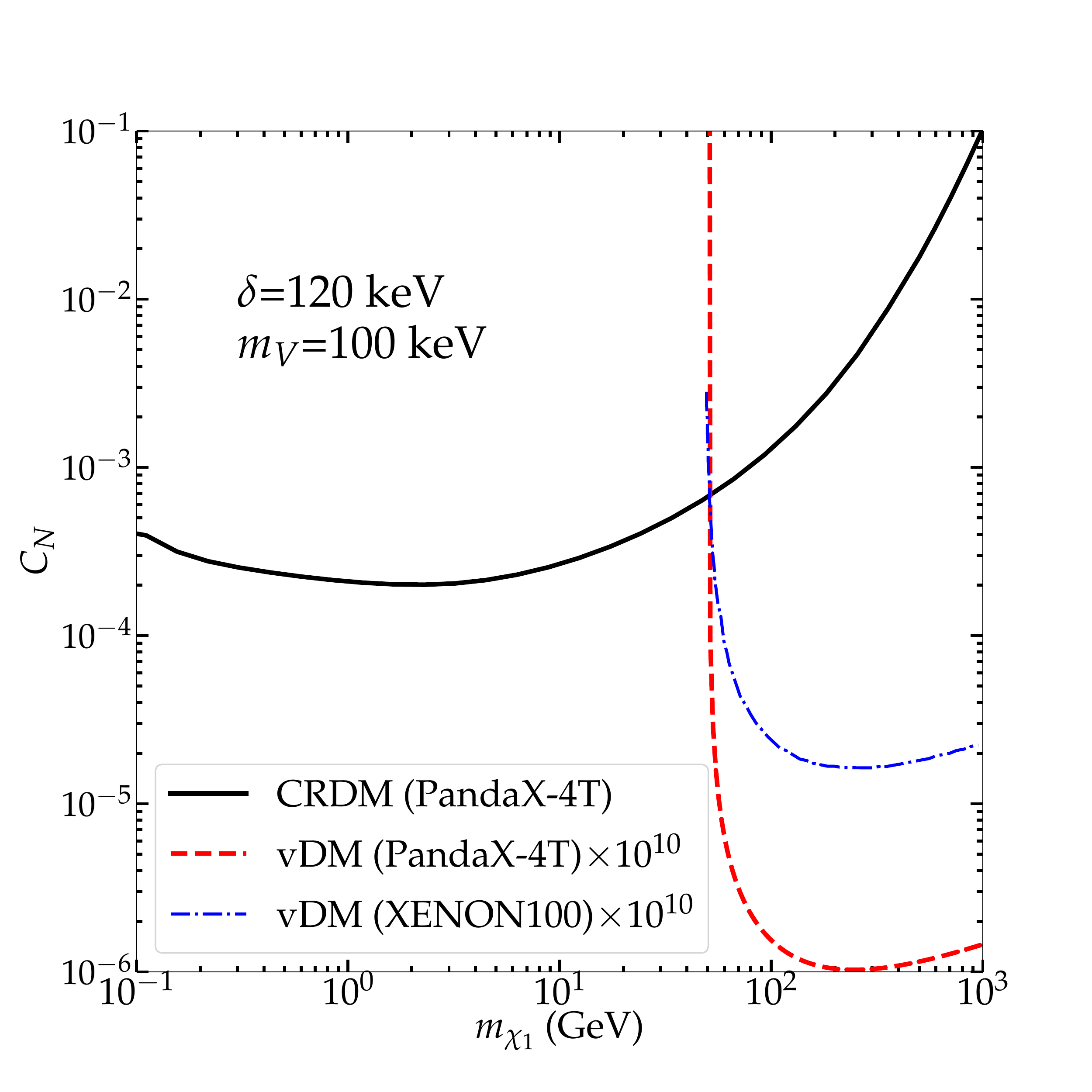}
\label{fig:vdm_crdm}
}
\subfloat[]{
\includegraphics[width=0.49\textwidth]{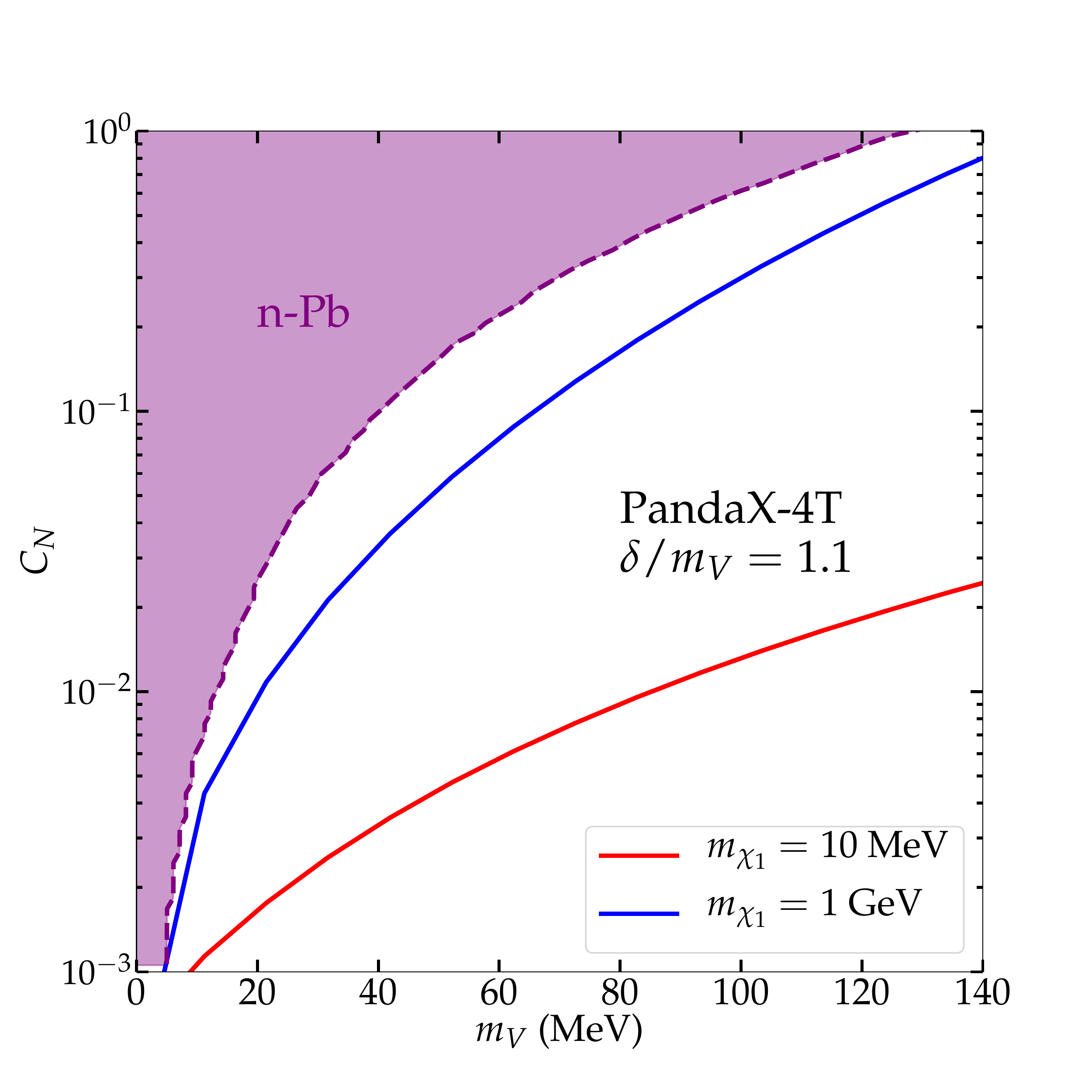}
\label{fig:mv_constraint}
}
\caption{The $95\%$ C.L. of PandaX-4T for both vDM (red dashed line) and CRDM (black solid line) in Fig.~\ref{fig:vdm_crdm}.   
As a comparison, we translate the XENON100 $95\%$ limit~\cite{XENON100:2011hxw} for vDM (blue dash-dotted line). 
We scale vDM limit with a factor $10^{10}$ in order to compare it with CRDM in a same range. 
The mass splitting and mediator mass are taken as $\delta=120 \kev$ and $m_V=100 \kev$. 
Fig.~\ref{fig:mv_constraint}: the comparison to the previous n-Pb scattering limits~\cite{Tulin:2014tya} 
in the ($m_V$, $C_N$) plane. 
The purple region is the exclusion of n-Pb scattering experiment while the PandaX-4T CRDM limits are presented 
by blue line ($1\gev$) and red line ($10\mev$). 
With $\delta=1.1\times m_{V}$, the limits of CRDM detection are more stringent than the previous n-Pb scattering experiment. }
\label{Fig:other_constraints}
\end{centering}
\end{figure*}

In Fig.~\ref{fig:vdm_crdm}, we compare the $95\%$ limits between vDM (dashed and dash-dotted) and CRDM (solid) scheme  
in ($C_N$, $m_{\chi_1}$) plane. 
Since the recoil energy of underground detectors can only reach $\delta \approx O(100\kev)$, 
we fix $\delta=120 \kev$ and $m_V=100 \kev$. 
Although the vDM limits are generally about 10 orders of magnitudes stronger than CRDM limits, 
they cannot probe the region $m_{\chi_1}\lesssim 50\gev$. 
Reversely, the CRDM limits can explore the parameter space for DM with a mass less than $50\gev$. 
We also translate the XENON100 limit from Ref.~\cite{XENON100:2011hxw} as a comparison. 
Similar constraints for inelastic vDM scheme can be found in Ref.~\cite{Schmidt-Hoberg:2009sgp,Chang:2008gd}.

In Fig.~\ref{fig:mv_constraint}, we show the upper limit of $C_N$ in a function of $m_V$ 
from n-Pb scattering (dashed purple) and our PandaX-4T CRDM limits (solid). 
We set $\delta=1.1\times m_{V}$ as a demonstration.
The most stringent constraint for $m_V\lesssim 100 \mev$ 
is derived from n-Pb scattering~\cite{Barbieri:1975xy,Leeb:1992qf,Tulin:2014tya}.
However, the CRDM limits can be still more stringent than the previous n-Pb scattering experiment, 
even if taking $m_{\chi_1}=1\gev$.

\end{document}